# Modelling ductile strain localization with evolutive stochastic rheologies

Andréa Tommasi (1)*, Felipe Sáez-Leiva (1), Lilou Zeller (1), Michel Peyret (1), Riad Hassani (2), Maurine Montagnat (3)

*(1) Géosciences Montpellier, CNRS & Université de Montpellier, France*
*(2) Géoazur, Observatoire de la Côte d'Azur, Université Côte d'Azur, France*
*(3) Université Grenoble-Alpes, CNRS, IRD, G-INP, IGE, France*
** Corresponding author: andrea.tommasi@umontpellier.fr*

**Abstract**

Plate Tectonics requires strain localization over the entire thickness of the plates. However, modelling strain localization in the deep sections of the plates, which deform by ductile processes, remains a challenge, prompting the use of ad hoc schemes to model plate boundaries. We posit that the bottleneck for self-consistent generation of ductile strain localization in geodynamical models is poor representation of the intrinsic mechanical heterogeneity of rocks, in particular at small scales. This prevents its effects from being accounted for at larger scales, notably emergent properties that arise during upscaling, like anisotropy. To model this heterogeneity and its evolution, we adopt a stochastic description of the rheology, which evolves in time and space as a function of the local work-rate. This approach enables to reproduce the full variety of responses observed in nature, from heterogeneous deformation at the local scale, but homogeneous at the system-scale, with or without softening, to spontaneous development of system-scale shear zones. It enables, thereby, the construction of regime diagrams for ductile strain localization using three adimensional parameters. These parameters describe the degree of heterogeneity and potential for evolution of the rheology, function of (1) the constitutive equation, which represents the active deformation processes, (2) the evolution law for the rock mechanical properties, (3) the initial properties, and (4) the energy input to the system. This approach also enables the prediction of the intensity of localization and the resulting system-scale anisotropic softening, paving the way for self-consistent modelling of plate boundaries in geodynamics.

**Highlights**

- Stochastic evolution of the rheology enables spontaneous viscous strain localization.
- Viscous localization stems from rheology variability controlled by the local work-rate.
- Local heterogeneities self-organize into system-scale shear zones.
- Strain localization produces anisotropic softening at the system-scale.
- Regime diagrams for strain localization for different viscous flow laws are similar.

## 1. Introduction

Shear zones - the geological expression of ductile (viscous) strain localization – are omnipresent in nature, from the mm to the planet scale. Their study reveals that rheological heterogeneity is the main cause of strain localization at all scales. This heterogeneity is ubiquitous at the rock scale. Most rocks are composed of minerals with different rheologies. Variations in mechanical behavior at this scale also arise from the strong viscoplastic anisotropy of most rock-forming minerals, like olivine or quartz, which results in contrasted strengths for crystals with different orientations relative to the imposed stresses (Doukhan and Trépied, 1985; Durham and Goetze, 1977). When grain-boundary processes are important, variations in grain size produce further heterogeneity in the mechanical behavior (Poirier, 1985). Both the mechanical behavior of the crystals that compose the rock and their spatial arrangement evolve in response to deformation. Fabric development and interconnection of weak phases are critical for enabling strain localization (Gerbi et al., 2016; Handy, 1994; Montési, 2013). This evolution, which involves processes like dynamic recrystallization, is largely controlled by the local stresses and strain rates. However, temperature, reactions, and fluids also play important roles (Braun et al., 1999; Holyoke and Tullis, 2006; Newman et al., 1999; Urai et al., 1986).

At the meter to kilometer scale, heterogeneity in the mechanical behavior arises from: (1) preexisting variations in lithology (Duretz et al., 2016; Neves et al. 2021), (2) the deformation itself, which produces shear heating and changes the rocks' structure, creating a fabric and, when this change is localized, shear zones (Drury et al., 1991; Montési, 2013; White et al., 1980), or (3) spatially heterogeneous fluid or melt distributions, which are often controlled by the deformation (Brown and Solar, 1998; Chardelin et al., 2024; Higgie and Tommasi, 2012; Katz et al., 2006; Tommasi et al., 1994). At even larger scales, rheological heterogeneity arises from spatial variations in the geotherm or mechanical anisotropy produced by mantle convection or the deformation history of the plates (Tommasi et al., 1995; Tommasi and Vauchez, 2015, 2001; Vilotte et al., 1982). Strain localization at the plates and mantle convection scales has also been proposed to arise from deformation-induced evolution of the rocks microstructure, notably grain size reduction (Bercovici and Ricard, 2014).

In summary, viscous strain localization, which produces the shear zones accommodating most of the ductile deformation on Earth, is controlled by feedbacks between physico-chemical processes occurring at a large range of spatio-temporal scales. This multi-scale and multi-physics character hindered self-consistent modelling of ductile strain localization in geodynamics so far. We can – albeit imperfectly – model microstructural evolution processes, like grain size reduction by dynamic recrystallization (Rozel et al., 2011) or development of layering (Dabrowski et al., 2012), and the associated softening. However, at larger scales these processes often result in the formation of shear zones and occur only within them. This point is critical, since the mechanical effect of microstructural evolution localized within a shear zone differs from that resulting from microstructural evolution affecting the entire rock volume. Yet, we currently lack tools to predict whether shear zones will spontaneously develop in response to deformation and, if so, quantify their effect on the mechanical behavior at larger scales. Previous works (Gardner et al., 2017; Gerbi et al., 2016; Handy, 1994; Meyer et al., 2017; Ruh et al., 2024) documented that both initial rheological heterogeneity and dynamic weakening are needed, but do not always suffice to produce interconnected weak domains, which localize shear and significantly modify the bulk rheology of a system. In this study, we aim to extend these results, obtained for specific configurations, by exploring a large parameter space to: (1) unravel the first-order parameters controlling the onset and stability of ductile strain localization at scales larger than the initial rheological heterogeneity characteristic length scale and, thereby, define a regime diagram for strain localization and (2) quantify the consequences of coupled microstructural evolution and strain localization on the bulk mechanical behavior.

## 2. An evolutive stochastic description of rheological heterogeneity

### *2.1 The model*

We address the heterogeneous nature of rock volumes and its evolution during deformation by implementing, in a finite-element code, a stochastic representation of the spatial variability of the mechanical properties of the medium, which evolves over time as a function of the local work-rate. Since rock volumes are mechanically heterogeneous due to a multitude of factors and since this heterogeneity evolves in response to multiple processes, we choose to solely model the resulting effect on the mechanical behavior. The formulation is kept as simple and general as possible, so that it may represent any structural evolution modifying the rheological parameters in a rock volume.

We consider a rock volume with an initially random distribution of mechanical heterogeneities. Instead of describing its rheology by an average of the local behaviors, we attribute to each mesh element a different rheological property randomly sampled from an initial, "undamaged" probability function ($P_0$) that describes the initial spatial variability in rheological properties within the modelled rock volume (Fig. 1, $t_0$). The rationale for describing the rheological properties by probability density functions and periodically perturbing the system is that most processes producing rheological heterogeneity are active at the sub-mesh scale. Stochasticity is required to describe their variability at the mesh scale.

As the volume starts to deform due to the imposed boundary conditions, heterogeneous stress and strain-rate fields develop (Fig. 1, $t_1$ top panel). Most processes producing rheological heterogeneity depend on deformation. The present model accounts for this by allowing the probability distributions P to evolve in a local manner, as a function of the local work-rate (Fig. 1, $t_1$ lower panel). If the local work-rate is higher or equal to a threshold value, the rock volume represented by the mesh element undergoes damage (a change in rheological properties, which produces softening). The rationale for regulating the rheology evolution by the local work-rate is that any structural evolution depends on a local balance between damage and healing rates, which is controlled by the work-rate (Austin and Evans, 2007; Bercovici and Ricard, 2005). Switch from damage to healing is therefore represented by a threshold on the work-rate.

Evolution of the rheological property field induces reorganization of the stress and strain-rate fields (Fig. 1, $t_{2,,n}$ top panel) and, by consequence, of the rheology field. Damage progresses up to a maximum $P_{dam}$ if the local work-rate remains higher than the threshold and previously damaged elements progressively heal if the local work-rate becomes lower than the threshold (Fig. 1, $t_{2,,n}$ lower panel). Over time, this feedback loop results in spatial reorganization of the rheological heterogeneity field. Coalescence of the damaged zones leads to a steady-state characterized by strain localization in a few system-scale shear zones.

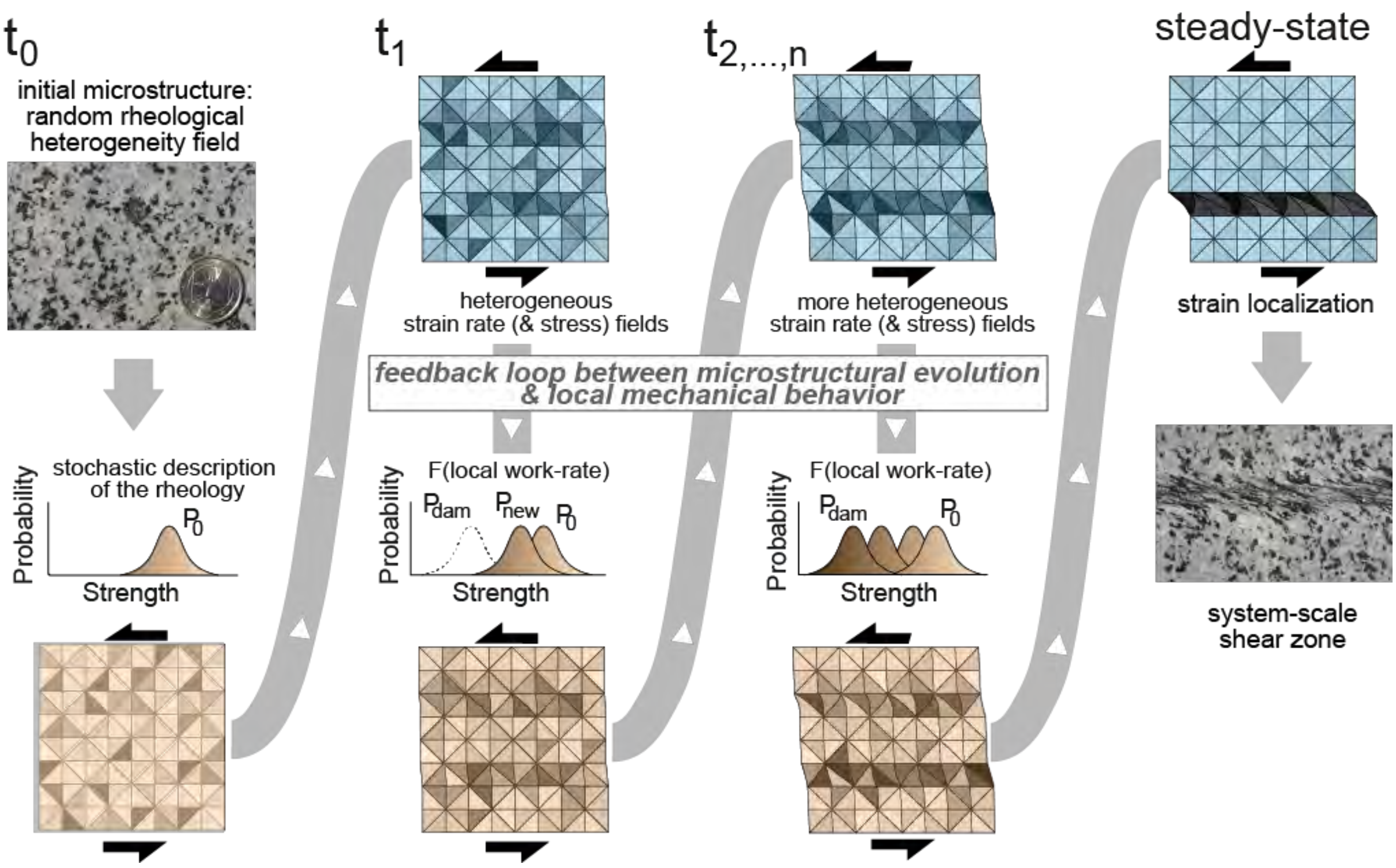

**Fig. 1:** Graphical summary of the approach proposed to self-consistently model system-scale strain localization arising from rheological heterogeneity. At time t0, rheological heterogeneity at the sub-mesh scale is modelled by randomly sampling a density probability function, which describes the initial spatial variability in a property controlling the material strength ($P_0$), schematically represented by variations in shades of brown in the plots. Spatial variation in strength produces heterogeneous strain rate (schematically represented by variations in shades of blue) and stress fields at time $t_1$. The material strength is then updated as a function of the local work-rate. If the latter is higher than a threshold, the mesh element is progressively damaged (its strength decreases). This is modeled by sampling the rheological property in another density probability function ($P_{new}$, which varies linearly between $P_0$ and $P_{dam}$) Otherwise, the mesh element either retains or progressively recovers its initial strength. Repetition over time ($t_{2,..,n}$) of this process results in spatial organization of the rheological heterogeneity field, leading to a steady-state characterized by strain localization in a small number (here, one) system-scale shear zones.

### *2.2 The numerical implementation*

This evolutive stochastic rheology is implemented in the finite-element code Adeli2024 (https://gitlab.oca.eu/hassani/adeli2024). This code is based on a Lagrangian discretization and uses a dynamic relaxation method to compute the quasi-static evolution of a rock volume behaving as an elastoplastic or viscoelastic medium.

In the present study, Maxwell viscoelastic rheologies are considered. Viscous creep is modelled by the semi-empirical constitutive equation:

$$\boldsymbol{D_v} = \gamma \exp\left(\frac{-Q}{RT}\left(1-\left(\frac{S_{eq}}{\tilde{\sigma}}\right)^{p}\right)^{q}\right) {S_{eq}}^{n-1}\boldsymbol{S} \qquad (1)$$

where $\boldsymbol{D_v}$ is the viscous strain-rate tensor, $S_{eq}$ is the Von Mises equivalent stress $\left(\frac{3}{2}\boldsymbol{S}:\boldsymbol{S}\right)^{1/2}$ with $\boldsymbol{S}$ being the deviatoric stress tensor, $R$ is the gas constant, and $T$ is the temperature. The material properties controlling the rheology are: $Q$ the activation energy in kJ mol$^{-1}$K$^{-1}$, $\gamma$ the fluidity in Pa$^{-n}$ s$^{-1}$, $\tilde{\sigma}$ the Peierls stress, and the constants *n, p,* and $q$ that modulate the dependence of the effective viscosity on $S_{eq}$. Depending on the choice of these parameters, this constitutive equation represents different deformation mechanisms. The full formulation models deformation by dislocation creep continuously from high to low stress conditions (Gouriet et al., 2019). The usual power law equation is retrieved by annulling the stress dependence within the exponential term ($q$=0). Deformation by diffusion creep, leading to a Newtonian behavior, can be modelled by further imposing n=1.

The evolution of the rheology (Fig. 2) is regulated by a work-rate threshold ($\dot{W}_{th}$), which determines, for each mesh cell, whether the instantaneous energy production enables damage (variation of the material property $P$ - in this study, $\tilde{\sigma}$ or $\gamma$ - producing softening). Otherwise the material heals, progressively retrieving its undamaged strength. Both damage and healing are implemented as linear evolutions of the mean of the material property $\bar{P}$ between that of initial undamaged rheology $P_0$ and that of the fully damaged rheology $P_{dam}$. $\bar{P}$ evolves according to the local strain, with independent scaling factors that allow for different kinetics of damage and healing ($K_{dam}$ and $K_{heal}$). These assumptions result in the following evolution equations:

$$\text{if } \dot{W}_{cell} > \dot{W}_{th} \text{ then } \begin{cases} \bar{P}_{new} = \bar{P}_{old} + \frac{\bar{P}_0 - \bar{P}_{dam}}{K_{dam}} \Delta J_2(\boldsymbol{\varepsilon}) \text{ if } \bar{P}_{new} \in [\bar{P}_{dam}; \bar{P}_0] \\ \bar{P}_{new} = \bar{P}_{dam} \quad \text{otherwise} \end{cases} \tag{2}$$

$$\text{else} \quad \begin{cases} \bar{P}_{new} = \bar{P}_{old} + \frac{\bar{P}_{dam} - \bar{P}_0}{K_{heal}} \Delta J_2(\boldsymbol{\varepsilon}) \text{ if } \bar{P}_{new} \in [\bar{P}_{dam}; \bar{P}_0] \\ \bar{P}_{new} = \bar{P}_0 \quad \text{otherwise} \end{cases}.$$

In a dynamic relaxation-based code, as the one used here, the rheology field cannot be updated at every timestep, as this increases the inertial forces. The material property P is therefore updated in all mesh cells at finite bulk strain intervals ($\Delta\left(J_2(\bar{\varepsilon})\right) = 0.001$) and, to ensure convergence, an additional condition on the residual of the inertial forces (< 0.05%) is imposed.

Both the initial and the evolving rheology are made stochastic by adding to the material property field $\bar{P}$ a Gaussian noise with spatial correlation (cf. Supplementary Material for the implementation). The Gaussian distribution was chosen, since it is the most general and easiest to implement. For future studies, different probability laws can be implemented. New noise is generated at every rheology update. This periodic perturbation of the rheological property field does not aim at representing a temporal evolution. It is intended as an equivalent representation of the sub-mesh heterogeneity of the rock. Correct sampling of the sub-mesh heterogeneity through multiple realizations per mesh cell, which is infeasible for the rheology in quasi-static mechanical models, would allow the emergence of stress and strain rate concentrations in every cell. Periodic perturbation of the rheology allows, over time, for stress and strain rate concentrations to appear across the entire modeled domain. It enables assessing their contribution to strain localization and testing the stability of the solutions.

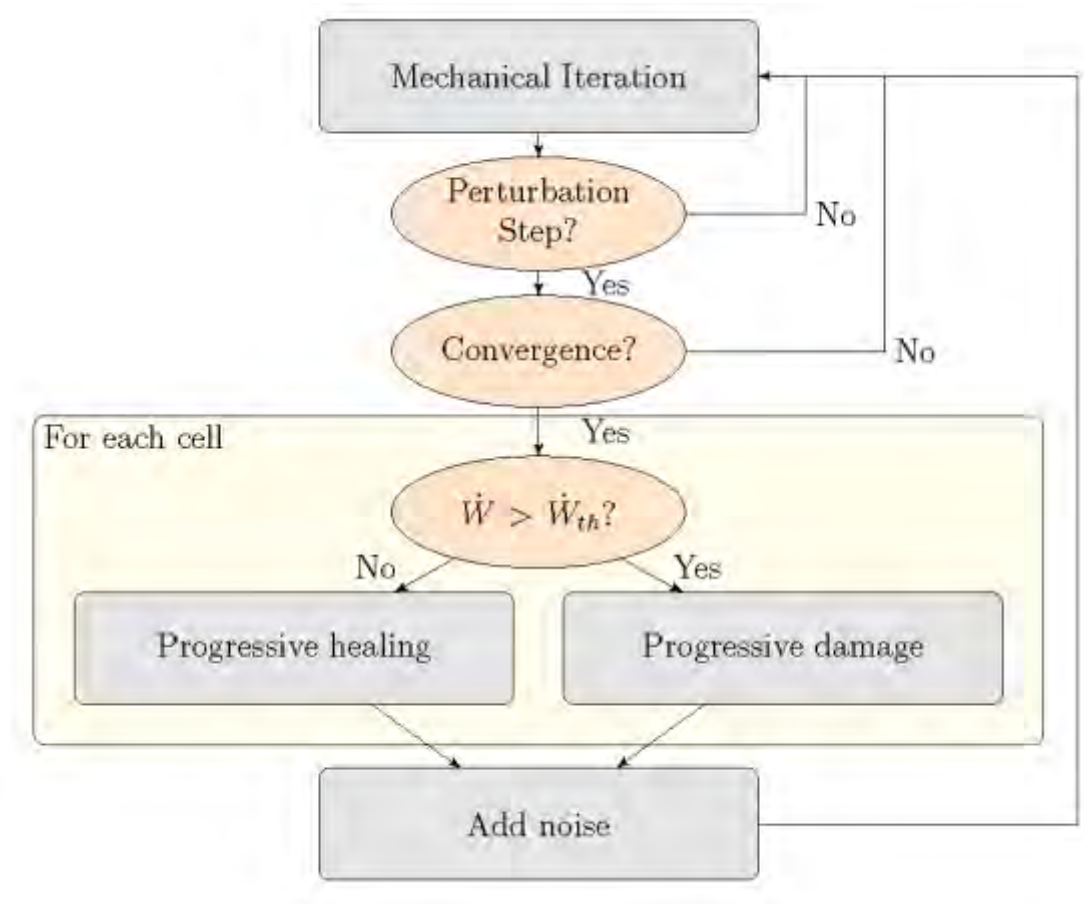

Fig. 2. Flowchart of the implementation of the rheology evolution.

### *2.3 The numerical experiments*

The numerical experiments have simple 2D-geometry and boundary conditions: a square domain subjected to shear at a constant rate of $10^{-14}$ $s^{-1}$ (Fig. 3). The temperature (1000 K) is constant both in space and time. The models have no intrinsic length scale, but the choices of uniform temperature and random initial rheological heterogeneity are better adapted to model the deformation at the rock (mm to m) scale.

The material parameters (Supplementary Materials Table S1) were adjusted to simulate the deformation of the lithospheric mantle, that is, olivine-rich rocks deforming by dislocation creep under moderate temperature conditions ($T/Tm$ = 0.7, where $Tm$ is the melting temperature). The rationale for this choice is twofold. First, except for extremely hot geotherms, the shallow lithospheric mantle is the strongest part of the plates, controlling their mechanical behavior. Second, analysis of naturally deformed samples and seismic anisotropy data indicate that the lithospheric mantle deforms dominantly by dislocation creep (Tommasi and Vauchez, 2015).

In most simulations, we impose the stochasticity on the Peierls stress ($P$=$\tilde{\sigma}$). Even within a monomineralic rock, $\tilde{\sigma}$ varies due to the crystals' mechanical anisotropy (Durham and Goetze, 1977), which results in spatially variable resistance to glide of dislocations. Dynamic recrystallization locally reduces $\tilde{\sigma}$ by producing new grains with low dislocation densities and more deformable orientations (Boissonneau et al., 2025; Urai et al., 1986). Given the highly non-linear strain rate dependence on $\tilde{\sigma}$ in the constitutive equation (eq. 1), a small decrease in $\tilde{\sigma}$ may produce marked softening. Healing - increase of the dislocation density and hence of $\tilde{\sigma}$, leading to hardening - is achieved by re-deformation of the recrystallized grains.

To test if the conclusions of the present study can be generalized to other (micro)structural evolution processes and deformation mechanisms, such as damage due to grain size reduction leading to increased contribution of grain boundary processes to deformation, simulations were also run for Newtonian and power law-type (n=3) constitutive equations with the stochasticity imposed on the fluidity ($P$=$\gamma$).

We conducted a parametric study focusing on five metrics, three of which appeared to have first-order control on strain localization: (1) the potential intensity of damage ($\pi_{dam}$), (2) the potential power for damage ($\pi_{\dot{W}}$), and (3) the spatial variability of the rheology ($\pi_{sto}$). These adimensional parameters and those used to quantify the strain localization and analyze its effect on the bulk mechanical behavior are presented in Table 1.

Table 1: Non-dimensional variables used to characterize the simulations

| | | Non-dimensional variable | Definition | Description |
|---|---|---|---|---|
| Inputs | | $\pi_{dam}$ | $\pi_{dam} = \frac{\lvert(mean(P_{dam}) - mean(P_0))\rvert}{mean(P_0)}$ | **Potential intensity of damage**, which is quantified based on the difference between the mean of the damaged probability distribution function and that of the initial, undamaged probability distribution function<br>$\pi_{dam}$ ranges from 0 (no damage) to infinite damage. |
| | | $\pi_{\dot{W}}$ | $\pi_{\dot{W}} = 1 - \frac{\dot{W}_{thr}}{\dot{W}_0}$ | **Potential power for damage** in the system, which is quantified by comparing the average work-rate in the system in its initial, non-damaged state ($\dot{W}_0$) and the threshold work-rate ($\dot{W}_{thr}$) for triggering damage. |
| | | $\pi_{sto}$ | $\pi_{sto} = \frac{stddev(P_0)}{mean(P_0)}$ | **Variability of the rheology**, which is quantified by the ratio between the standard deviation and the mean of the probability distribution function. In the present study, this ratio is equal for the initial and damaged probability distribution functions. |
| Outputs | Intensity of strain localization | ${V_{loc}}^*$ | ${V_{loc}}^* = \frac{V_{loc}}{V}$ | **Volume fraction of the system accommodating localized deformation**. $V_{loc}$ is defined as the sum of the volumes of the mesh elements that accommodate the upper 50% of the bulk Von Mises equivalent strain rate. V is the total volume of the simulation box. In 2D simulations, as in this study, area instead of volume is used to define $V_{loc}$*.<br>${V_{loc}}^*$ ranges from 0.5 (no strain localization) to 0 (infinite strain localization in a plane of null width, which is only achieved for plastic deformation) |
| | | ${D_{loc}}^*$ | ${D_{loc}}^* = \frac{\frac{1}{Vloc}\int_{Vloc} D_{eq} dv}{\frac{1}{V}\int_v D_{eq} dv}$ | **Strain rate enhancement**, defined as the ratio of the average Von Mises equivalent strain rate in $V_{loc}$ relative to the bulk Von Mises equivalent strain rate in the system, with $D_{eq} = \left(\frac{2}{3}\boldsymbol{D}:\boldsymbol{D}\right)^{1/2}$<br>${D_{loc}}^*$ ranges from 1 (no strain localization) to infinite (infinite strain localization) |
| | | $\pi_\eta$ | $\pi_\eta = \log_{10} \frac{\frac{1}{V_{Nloc}}\int_{V_{Nloc}} \eta dV}{\frac{1}{V_{loc}}\int_{V_{loc}} \eta dV}$ | **Contrast (logscale) in mean effective viscosity** in $V_{loc}$ relative to that in the background ($V_{Nloc}$=V-$V_{loc}$)<br>$\pi_\eta$ ranges from 0 (no strain localization) to infinite (infinite strain localization) |
| | Effect on the bulk mechanical behavior | $\pi_{soft}$ | $\pi_{soft} = 1 - \frac{\int_v {S_{eq}}_{ss} dv}{\int_v {S_{eq}}_0 dv}$ | **Effective bulk softening**, which, in the present models with imposed velocity conditions, is calculated based on the ratio between the final steady-state Von Mises equivalent bulk stress ($S_{eq}$) relative to that of the initial, non-localized state<br>$\pi_{soft}$ ranges from 0 (no softening) to 1 (infinite softening) |
| | Steady-state is defined by a variability of ${V_{loc}}^*$ , ${D_{loc}}^*$, $\pi_\eta$, and $\pi_{soft}$ (estimated over the last five perturbations) smaller than the uncertainty due to stochasticity, which is estimated based on ensembles of simulations with different random initializations. | | | |

We also tested the sensitivity of the solutions relative to the damage and healing kinetics, the noise spatial correlation, and mesh refinement. The reference and ranges of tested values are given in Table 2. To probe the effect of different random initializations of the system, for certain sets of parameters, ensembles of 20 simulations with different seeds were run.

Table 2. Parametric study: reference values and explored ranges

| Parameter | Reference value | Tested range | Associated adimensional parameter range |
|---|---|---|---|
| Peierls stress ($\tilde{\sigma}_0$), GPa | undamaged Peierls stress ($\tilde{\sigma}_0$) = 2 | damaged Peierls stress ($\tilde{\sigma}_{dam}$) = [0.75, 1.0, 1.25, 1.5, 1.75] | Potential intensity of damage ($\pi_{dam}$) = [0.125, 0.25, 0.5, 0.625] |
| Work-rate threshold for damage ($\dot{W}_{thr}$), Pa/s | average work-rate in the system in its initial, non-damaged state ($\overline{\dot{W}_0}$) = 2.55e6 | Work-rate threshold for damage ($\dot{W}_{thr}$) = [$0.1\times \overline{\dot{W}_0}$ , $2 \times \overline{\dot{W}_0}$] | Potential power for damage ($\pi_{\psi}$) = [-1, 0.9] |
| Peierls stress standard deviation | $0.25 \times \tilde{\sigma}$ | [0.001,0.125,0.25,0.375] * $\tilde{\sigma}$ | Spatial variability of the rheology ($\pi_{sto}$) = [0.001, 0.125 ,0.25, 0.375] |
| Damage and healing kinetics ($K_{dam}$ - $K_{heal}$) | 0.01-0.01, which implies damage and healing kinetics 10 times faster than the strain rate | [0.005 - 0.005, 0.01 - 0.01, 0.02 - 0.02, 0.01 - 0.5, 0.01 - 1] | |
| Spatial correlation of the noise | $0.005 \times$ model length | [0.0025, 0.005, 0.01, 0.025] | |
| Mesh refinement | mesh cell = 0.005 x model length | [0.00025, 0.01] | |

## 3. From random local rheological heterogeneity to system-scale shear zones

The present numerical experiments confirm that strain localization at the system-scale may, but does not necessarily, arise self-consistently from local rheological heterogeneity. They show that the onset of strain localization at scales larger than those of the initial heterogeneities depends on the interplay between the evolutions of the rheology and work-rate distribution. When it arises, strain localization is stable, leading to a new steady-state of the system.

The possible evolution paths are illustrated in Figure 3, which presents experiments in which stochasticity is imposed on the Peierls stress. These paths encompass the entire spectrum of behaviors observed in nature and experiments: from strain localization in shear zones of variable thicknesses (cf. references in the introduction and Barnhoorn et al., 2005; Holyoke et al., 2014) to heterogeneous deformation at the grain scale - yet homogeneous at the sample scale - sometimes accompanied by bulk softening, usually observed in torsion experiments in monomineralic aggregates (Bystricky et al., 1999; Hansen et al., 2014; Pieri et al., 1999) or rocks deformed at high temperature conditions.

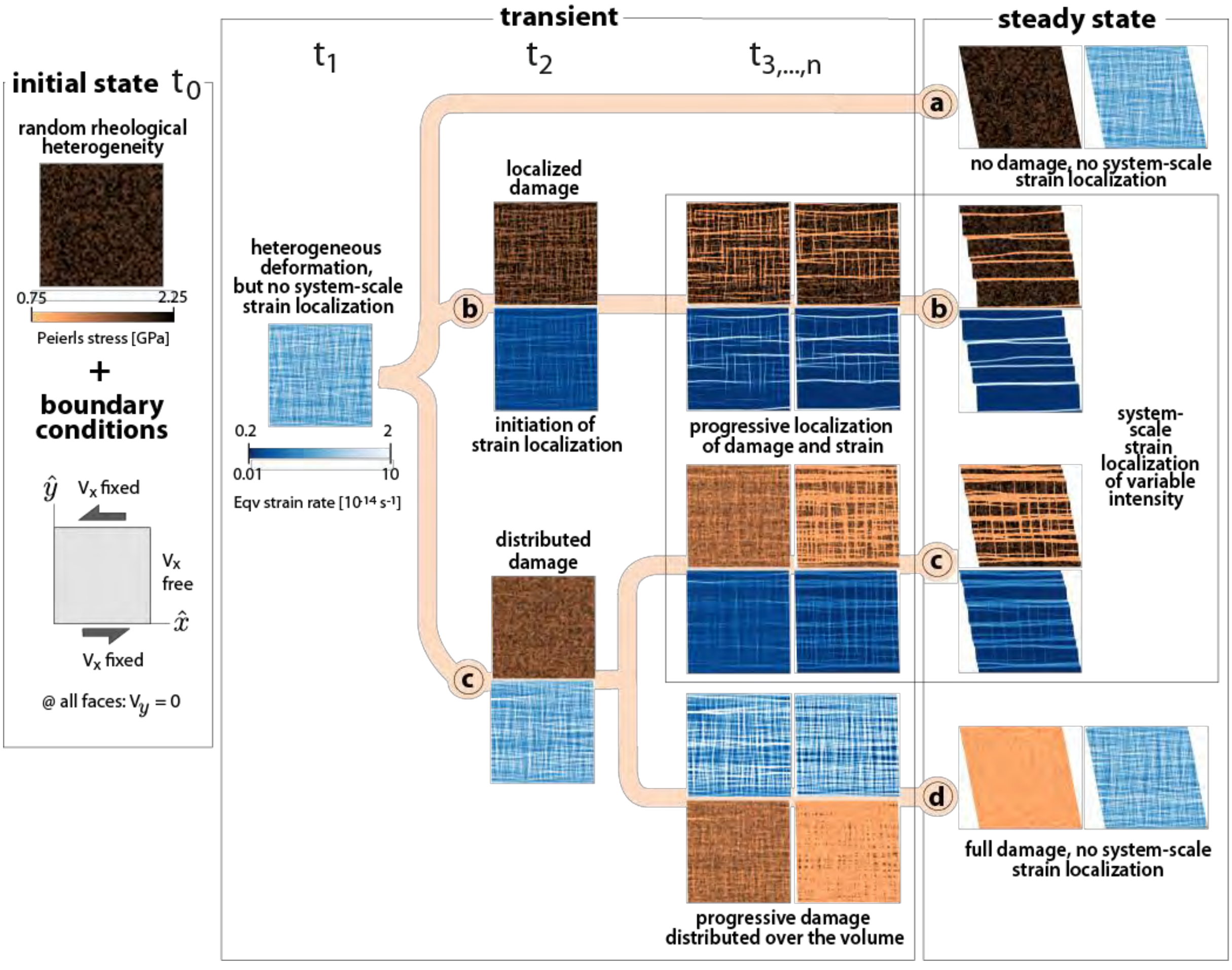

Fig. 3: The possible evolution paths for a viscous medium with an initially random rheological heterogeneity field, which can evolve through damage and healing as a function of the local work-rate, subjected to simple constant boundary conditions (here, simple shear at a bulk strain rate of $10^{-14}$s$^{-1}$). For each path, the evolution of the material property controlling the rheology and varying due to damage and healing (here, the Peierls stress, in shades of orange) and the strain rate (in shades of blue) fields are presented. For times higher than $t_1$, the strain rate colorbars differ between the simulations with strain localization (b & c : $10^{-16}$ - $10^{-13}$s$^{-1}$) and those without strain localization (a & d : 2x$10^{-15}$ - 2x$10^{-14}$s$^{-1}$).

The initial stage is similar for all paths (Fig. 3, $t_1$). Random spatial variation in the rock mechanical properties produces heterogeneous strain rate and stress fields at the scale of the heterogeneities. The stress field (not shown in Fig. 3) is directly correlated to the rheological heterogeneity field. The strain rate field is characterized by weak strain localization within discontinuous, asymmetric lenses preferentially oriented at 45° to the maximum and minimum compressive stresses. Strain rates within these incipient shear zones are less than twice that of the background. Their width is controlled by the characteristic length scale of the rheological heterogeneity field. Their length and spacing are controlled by the convolution of the strain rate perturbations produced by all rheological heterogeneities in the system, being 2-10 times the characteristic length scale of the rheological heterogeneity field.

In absence of damage (no evolution of the probability distribution function of the property controlling the rheology), the strain rate field evolves as a function of the successive perturbations of the rheological property field, since unique stable strain rate and stress fields are associated with each rheology field.  In the present simulations, the boundary conditions favor the development of high strain rate lenses forming an anastomozed pattern with a dominant trend parallel to the imposed shear (Fig. 3a). However, further strain localization into shear zones cross-cutting the entire simulation domain (system-scale strain localization) never arises.

Strain localization can develop at the system-scale if the stochastically-described material property evolves following the damage/healing process described in the previous section. After a transient (Fig. 3b), in which some of the initial strain localization lenses lengthen and coalesce, widening or not, while the others progressively disappear, the system reaches a new steady-state. The evolution during the transient is controlled by the feedbacks between the perturbations in the work-rate field produced by all rheological heterogeneities in the system. The steady-state is characterized by shear zones that crosscut the entire simulation domain (Fig. 3b). The intensity of strain localization is inversely correlated to the possibility of damage to occur. When only small fractions of the modeled domain can damage (display a work-rate higher than the threshold), few shear zones, which strongly localize strain, develop (Fig. 3b). When large fractions of the system can damage, more shear zones develop and strain localization is weaker (Fig. 3c).

However, damage does not result systematically in system-scale strain localization. When the bulk work-rate is higher than the damage threshold, large fractions or even the entire modeled domain may be damaged. The system may then evolve towards a steady-state geometrically similar to the non-damaged one, but characterized by a lower bulk strength (Fig. 3d).

Analysis of ensembles of simulations with different random initializations (Fig. 4) shows that the number, location, and thickness of the individual shear zones vary, but the statistical properties of the system, like the volume fraction accommodating the localized deformation ($V_{loc}$*, Fig. 4a), the contrast in strain rate between the shear zones and the background ($D_{loc}$*), and the bulk softening ($\pi_{soft}$), converge towards steady values with standard deviations <10%. The mean values at steady-state depend, at first-order, on the reduction of effective viscosity produced by damage, which is controlled by the potential intensity of damage ($\pi_{dam}$) and, secondarily, on the potential power for damage ($\pi_{\dot{W}}$) and the variability of the rheology ($\pi_{sto}$).

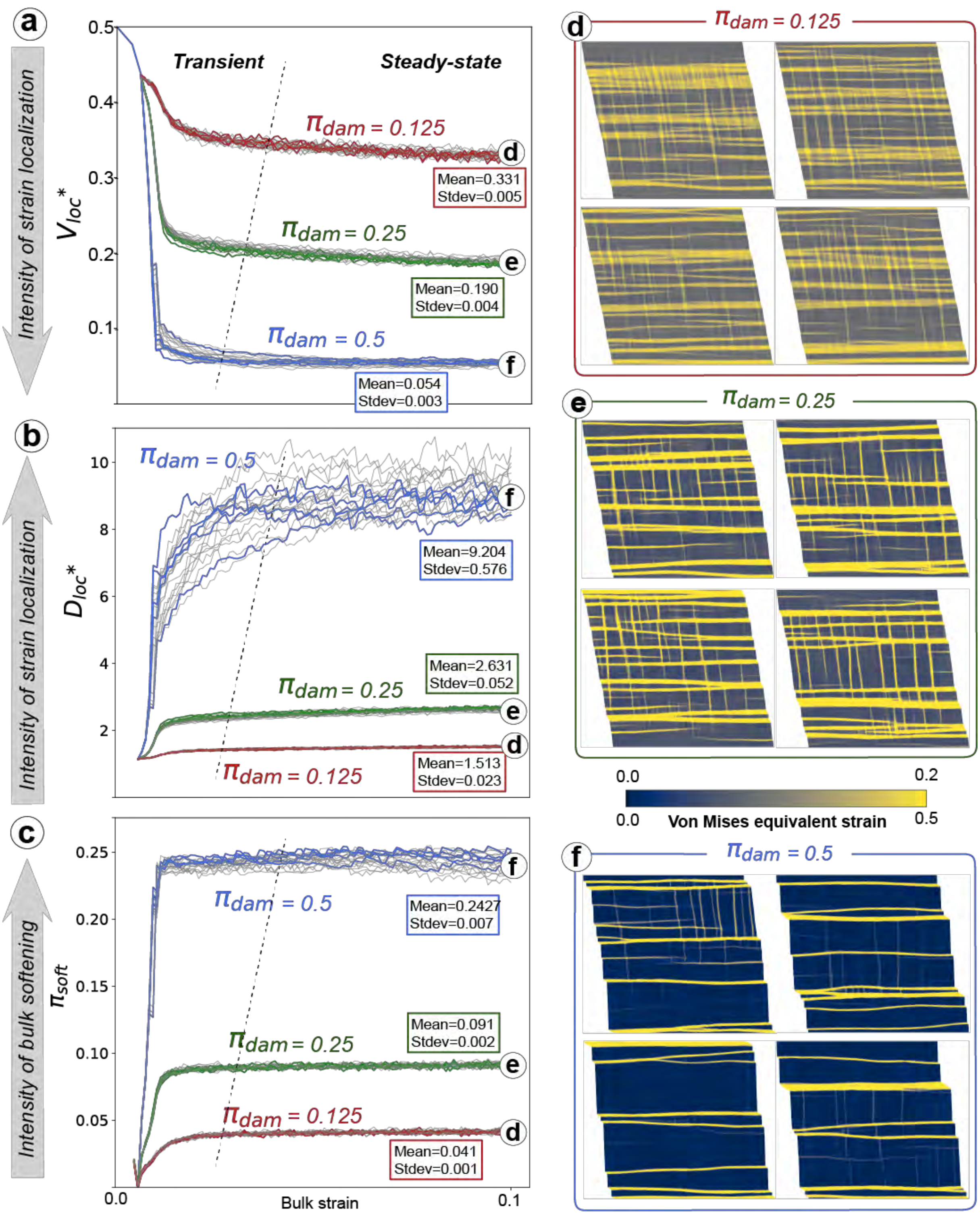

Fig. 4. Evolution of the system towards steady-state strain localization, quantified by (a) the volume fraction of the system accommodating the localized deformation ($V_{loc}$*), which decreases with increasing strain localization, (b) the average strain rate in the shear zones normalized by the bulk strain rate ($D_{loc}$*), and (c) the bulk softening ($\pi_{soft}$) in three ensembles of 20 simulations with different random initializations and increasing potential damage intensity ($\pi_{dam}$). All other parameters are similar. (d), (e), and (f) Steady-state finite strain ($J_2(\varepsilon)$) fields for four different realizations of each ensemble, illustrating that different random initializations result in different number, location, and widths of individual shear zones (in yellow), despite the similar bulk (average) behavior of the system, which is mainly controlled by $\pi_{dam}$. Note that the strain range in panels (d) and (e): [0,0.2], differs from that in panel (f): [0,0.5].

## 4. Regime diagrams for viscous strain localization

Analysis of the ensemble of numerical simulations highlights that development of system-scale strain localization during deformation accommodated by viscous processes requires: (1) initial heterogeneity in the rheological behavior, (2) evolution of the rock structure (damage/healing) producing spatial variations in the rheology controlled by the mechanical energy dissipation field, and (3) a bulk mechanical energy dissipation in the system within a range of work-rate threshold for damage, which increases with increasing rheology variability. We may thus construct regime diagrams for ductile strain localization (Fig. 5) as a function of the potential power for damage ($\pi_{\dot{w}}$), the potential intensity of damage ($\pi_{dam}$), and the initial variability in the material properties ($\pi_{sto}$, cf. Table 1 for definitions). The first and third non-dimensional numbers control whether damage and healing may be triggered. The second regulates, via the constitutive equation, the intensity of mechanical behavior changes - softening or hardening - associated with damage or healing.

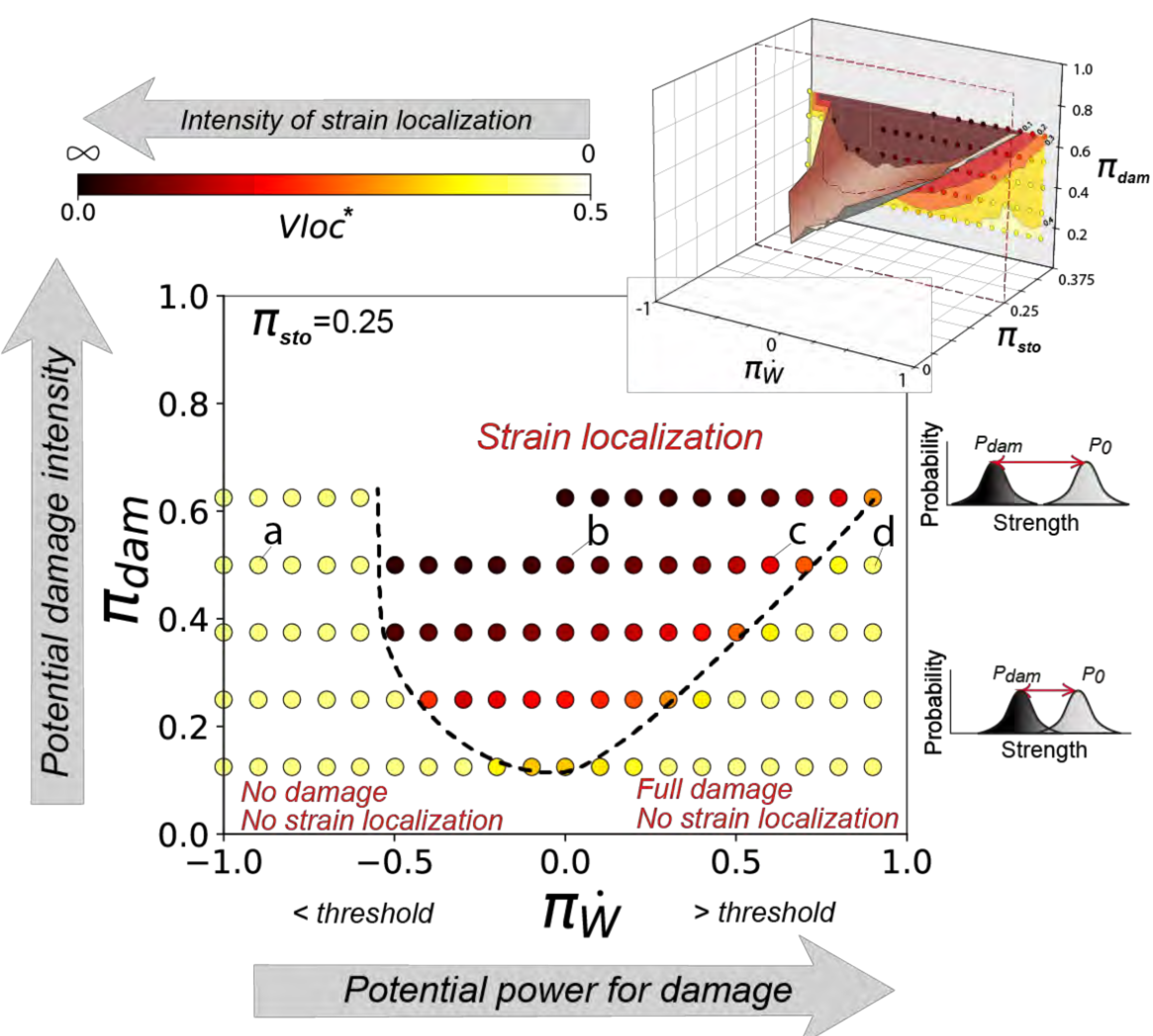

Fig. 5. Regime diagram for viscous strain localization during deformation by dislocation creep at moderate temperature (stochasticity imposed on the Peierls stress) as a function of the potential power for damage ($\pi_{\dot{w}}$), the potential intensity of damage ($\pi_{dam}$), and the spatial variability of the rheology ($\pi_{sto}$). Insert displays the $V_{loc}$* = 0.1 envelope in the full 3D regime diagram. The main figure is the $\pi_{\dot{w}} - \pi_{dam}$ section at $\pi_{sto}$ =0.25 with the intensity of strain localization at steady state in each simulation indicated by the color of the symbol. Darker colors denote higher strain localization. Missing points are simulations in which extreme strain localization produced mesh degeneration. $\pi_{\dot{w}} - \pi_{dam}$ sections at different $\pi_{sto}$ values are presented in Supp. Mat. Fig. S1.

The regime diagram shows that system-scale strain localization is the stable solution within a 3D domain with an asymmetric V-shape on the $\pi_{\dot{w}} - \pi_{dam}$ plane, which widens with increasing $\pi_{sto}$. In this domain, strain localization at steady-state is quantified by $V_{loc}{}^{*}$, the volume fraction of the system accommodating half of the total strain rate, which decreases with increasing strain localization (Figs. 5 and 6a) and by $D_{loc}{}^{*}$, the ratio between the strain rate within $V_{loc}{}^{*}$ and the bulk strain rate, which increases with increasing strain localization (Fig. 6b).

The strain localization domain is surrounded by two domains in which the deformation is heterogeneous at the microscale, but statistically homogeneous at the system-scale ($V_{loc}{}^{*} = 0.5$ and $D_{loc}{}^{*} = 1$, Fig. 6a). The two non-localizing domains differ, though, in bulk mechanical behavior (Fig. 6c). In one, significant damage is never achieved, because the local work-rate production is too low to produce damage. By consequence, its bulk mechanical behavior remains constant ($\pi_{soft}$=0 in Fig. 6c). In the other, the threshold for damage is surpassed everywhere, resulting in full-system damage and in the maximum bulk softening that may be achieved for a given $\pi_{dam}$. The bulk mechanical behavior in either absence of damage or full damage cases is similar to that of a deterministic model with a homogeneous rheology defined by the mean of $P_0$ ($\pi_{soft} = 0$) or $P_{dam}$ ($\pi_{soft}$ max, F($\pi_{dam}$)), respectively.

Increasing the potential damage intensity ($\pi_{dam}$) and, by consequence, the potential reduction of the effective viscosity, enhances the strain localization probability and intensity ($V_{loc}{}^*$ decreases in Fig. 5). For high potential damage, strong strain localization is the stable solution for a large range of $\pi_{\psi}$; this range increases with increasing $\pi_{sto}$. At steady-state, the imposed strain rate is almost entirely accommodated in a few shear zones (Fig. 3), which may occupy <5% of the total volume and deform more than 10 times faster than the imposed strain rates ($V_{loc}* < 0.05$ and $D_{loc}* > 10$, Figs. 4 and 6a,b). Strain rates in these shear zones are up to three orders of magnitude higher than the background (Fig. 3b).

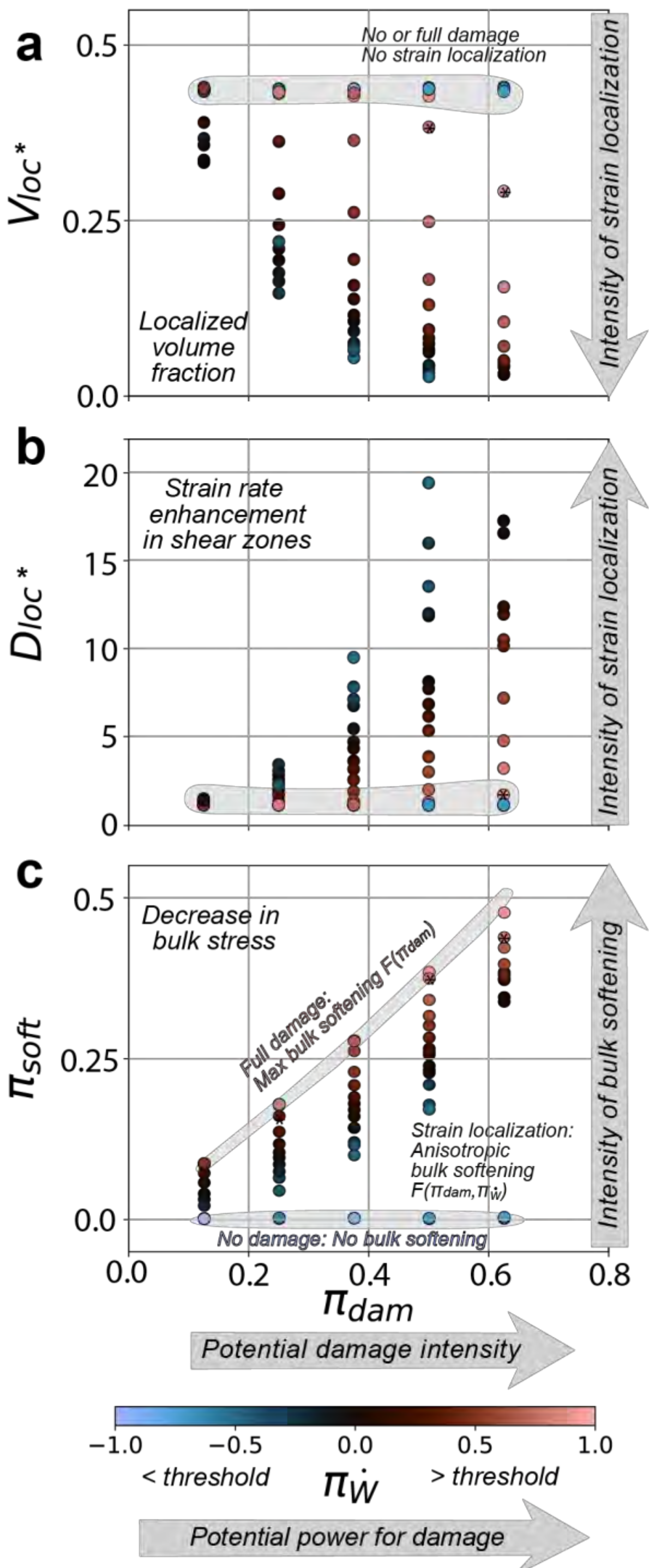

Fig. 6. Intensity of strain localization at steady-state described by (a) the localized volume fraction ($V_{loc}{}^{*}$) and (b) the ratio of the mean strain rate in the shear zones relative to the bulk strain rate in the system ($D_{loc}{}^{*}$). (c) Bulk softening, characterized by the decrease in bulk stress relative to the initial, non-localized state ($\pi_{soft}$) for all simulations in the regime diagram section in Fig. 5. Symbols are colored as a function of the potential power for damage ($\pi_{\dot{w}}$). * mark runs that have not achieved steady-state at the end of the simulation (bulk strain of 0.1).

For low potential damage, strain localization is only observed when the average work-rate production in the initial, non-damaged system is similar to the damage threshold work-rate ($\pi_{\psi} \sim 0$; Fig. 5). This is particularly true when the initial heterogeneity is weak (low $\pi_{sto}$). Low potential damage produces weak rheological contrasts. By consequence, multiple wide shear zones form and evolve into steady-state. These shear zones occupy a large fraction of the system (up to 30%) and are characterized by low strain rate enhancements (factor <10) relative to the background (Figs. 4 and 6a,b).

For a given potential damage, the duration of the transient and the strain localization intensity at steady-state depend on the potential power for damage ($\pi_{\dot{W}}$). Within the strain localization domain, an initial bulk work-rate lower than the threshold ($\pi_{\dot{W}} < 0$) results in rare damage events, development of a small number of shear zones, which strongly localize strain (Figs. 3b, 6a,b), and short transients (Supplementary Material Fig. S5 and S6). Initial bulk work-rates higher than the damage threshold ($\pi_{\dot{W}} > 0$) lead to longer transients and weaker strain localization (Fig. 3d). The higher the 'excess' work-rate production in the system, the longer is the transient (Supplementary Material Figs. S5 and S6) and the weaker the intensity of localization at steady-state (lower $V_{loc}{}^*$ and higher $D_{loc}{}^*$ in Fig. 6a,b). Since the intensity of the bulk softening is proportional to the damaged volume, at constant $\pi_{dam}$ and $\pi_{sto}$, higher $\pi_{\dot{W}}$ results in stronger bulk softening (higher $\pi_{soft}$ in Fig. 6c).

Since strain localization for low $\pi_{\dot{W}}$ depends on rare damage events, longer simulation times or larger domains enhance the probability of these events and hence enlarge the strain localization domain at the expenses of the non-damage one. Ensembles of 50 simulations for cases in the non-damage domain close to the limit with the strain localization show indeed strong strain localization in one or two cases. The regime diagram was constructed based on the dominant behavior of the system at the studied conditions, ignoring these extreme behavior cases.

The other tested parameters have second-order effects. Regime diagrams for different spatial correlations of the noise and kinetics of damage and healing display similar V-shape strain localization domains with similar volumes (Table 3). Only the location of the domains' limits in the regime diagram and the stability of the solutions change slightly (cf. Supplementary Materials Figs. S1-S7). Noteworthy results are: (1) the evolution and steady-state bulk properties are almost independent on the initial rheological heterogeneity field, (2) the damage and healing kinetics control the transient, but have a weak effect on the steady-state. Resolution tests document that viscous strain localization is not mesh-dependent (Supp. Mat. Fig. S3).

Table 3: Effect of tested parameters on strain localization at steady-state, estimated by the difference in the area fractions with strong and weak strain localization in each regime diagram relative to that of the reference simulation.

| Parameter | Tested value | Area with strong strain localization $V_{loc}$* ≤ 0.1 | Area with weak strain localization $V_{loc}$* ≥ 0.4 |
|---|---|---|---|
| Spatial correlation (ref =0.005) | 0.025 | +0.04 | -0.06 |
| | 0.0025 | -0.03 | +0.08 |
| Standard deviation (ref = 0.25 × mean) | 0.001 × mean | -0.22 | +0.33 |
| | 0.375 × mean | +0.09 | -0.21 |
| Kinetics of damage and healing $K_{dam}$ – $K_{heal}$ (ref = 0.01- 0.01) | 0.005 – 0.005 | 0 | -0.06 |
| | 0.02 – 0.02 | -0.06 | +0.10 |
| | 0.01 - 1 | -0.03 | -0.02 |
| | 0.01 – 0,5 | -0.04 | +0.01 |

Regime diagrams for strain localization obtained for Newtonian and "power-law" materials with variable fluidity (Fig. 7a) have similar shapes as those resulting from variability in the Peierls stress. However, since the strain rate depends linearly on the fluidity, much higher $\pi_{dam}$ values are needed to produce the same intensity of strain localization ($V_{loc}$*). The strong contrasts in fluidity required to produce strain localization in the stochastic fluidity simulations led to unstable simulations, which alternate full-volume damage and softening or show extreme strain localization triggered from the boundaries, in particular for Newtonian materials (gray domain in Fig. 7a).

Bulk softening curves as a function of $\pi_{\dot{w}}$ for the three rheologies also have similar shapes (Fig. 7b), but the intensity of softening ($\pi_{soft}$) and the mean effective viscosity contrast between the shear zones and the surrounding medium ($\pi_{\eta}$) at constant $\pi_{dam}$ and $\pi_{sto}$ depend strongly on the non-linearity of the flow law. Decreasing $\tilde{\sigma}$ by a factor two results in effective viscosity contrasts of two orders of magnitude and a reduction of the bulk strength by a factor two. Much higher $\gamma$ variations are required to produce similar behaviors. However, the possible ranges of $\pi_{dam}$ for $\tilde{\sigma}$ and $\gamma$ differ. For the flow laws used in this study, $\pi_{dam}$=1 represents an infinite damage if the evolution is imposed on $\tilde{\sigma}$, but an increase of $\gamma$ by a factor two. Grain size reduction may increase $\gamma$ by more than six orders of magnitude.

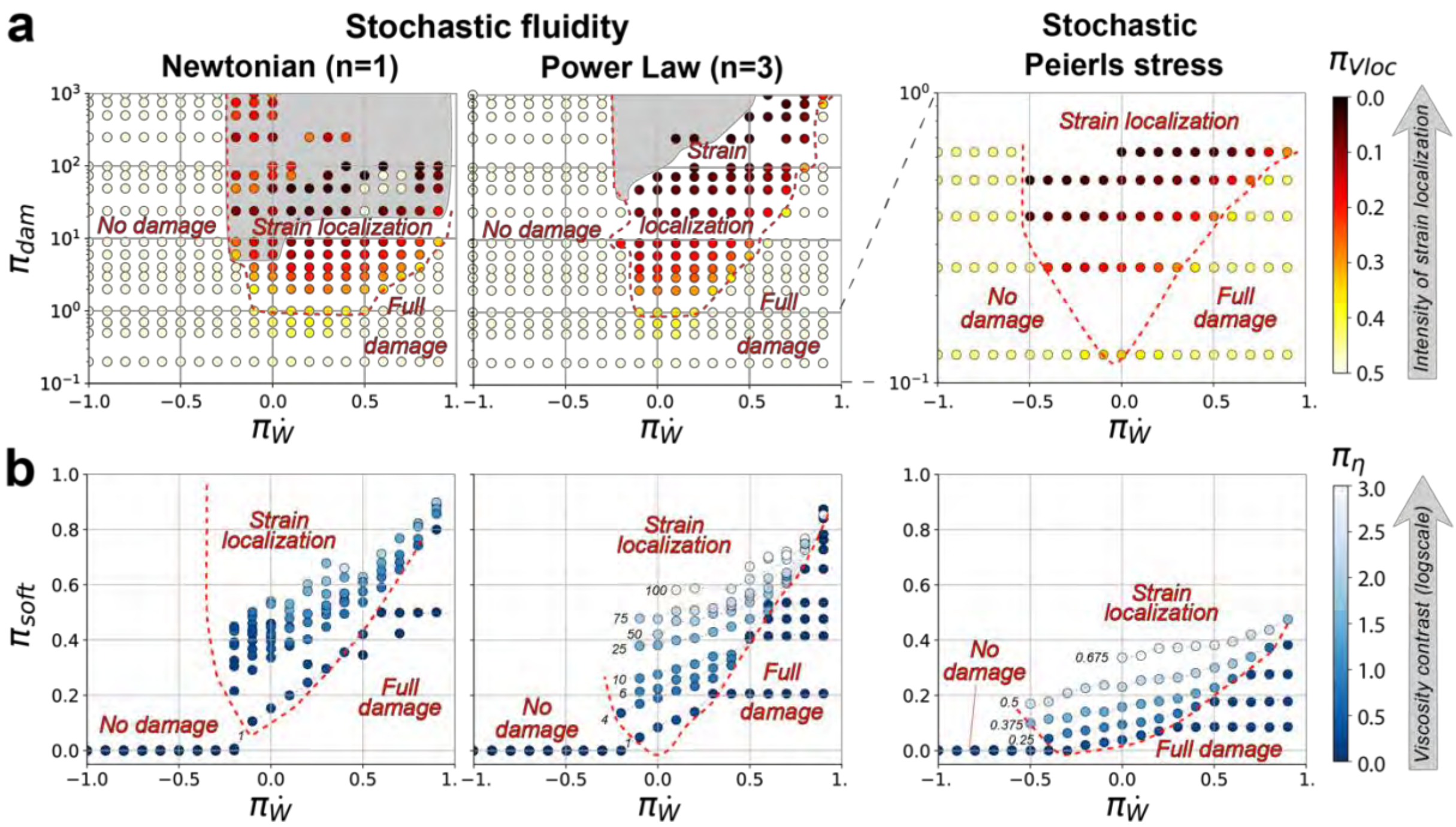

Fig. 7. Comparison of (a) the $\pi_{\dot{W}} - \pi_{dam}$ sections at $\pi_{sto}$=0.25 of the regime diagram for viscous strain localization and (b) plots of the bulk softening at steady state ($\pi_{soft}$) as a function of the potential power for damage ($\pi_{\dot{W}}$) in Newtonian or power-law materials where stochasticity and evolution are imposed on the fluidity to those obtained in the reference model, where stochasticity and evolution are imposed on the Peierls stress. In (a) the different experiments are colored as a function of the intensity of strain localization ($V_{loc}{}^{*}$), whereas in (b) they are colored as a function of the logarithm of the mean viscosity contrast between the shear zones and the surrounding medium ($\pi_{\eta}$). Note the different ranges of $\pi_{dam}$ in (a). The domain shaded in grey in the plots for Newtonian and Power Law materials in (a) represents unstable simulations. Thin dashed lines in (b) materialize isocontours of potential damage intensity ($\pi_{dam}$).

## 5. Discussion

### *6.1 Model strengths and limitations*

By explicitly accounting for the evolution in both space and time of the rheology, the present approach enables modelling self-consistent generation of system-scale ductile strain localization from local random rheological heterogeneity. The present work builds on and generalizes previous works (Gardner et al., 2017; Gerbi et al., 2016; Handy, 1994; Meyer et al., 2017; Ruh et al., 2024) that documented that both initial rheological heterogeneity and dynamic weakening are necessary, but do not suffice to produce interconnected weak domains, which localize shear and significantly modify the bulk rheology of a system. The present results are consistent with the predictions of these models for the onset of strain localization, its intensity, and associated bulk softening. These studies explored, however, a limited space of rheology contrasts and geometrical configurations.

By adopting a stochastic evolutive description of the rheology, coupling its evolution to the local mechanical dissipation rate, and exploring a large parameter space, the present models achieve enough degrees of freedom to capture the first-order parameters controlling the development or not of strain localization, as well as its intensity and consequences on the rheology at the system scale. These parameters are: the initial microstructure (average properties and variability), the maximum admissible damage and the associated softening, and the boundary conditions, which describe the energy input to the system and, thereby, influence the rheology evolution. They were combined to construct a regime diagram for strain localization based on three adimensional parameters. This simplified, but explicit description of the spatial and time variability of the rheology is essential for self-consistent triggering of strain localization, which arises from a feedback between spatial variations in the mechanical fields induced by rheological heterogeneity and the evolution of the rheology (Fig. 1).

The inherent spatial heterogeneity of the rocks mechanical behavior is represented by $\pi_{sto}$. The potential damage intensity ($\pi_{dam}$) accounts implicitly for the rocks' initial properties and the processes changing these properties and, by consequence, the rheology. In nature this evolution involves multiple, complex physico-chemical processes, which occur at a variety of spatial and time scales, but mostly at the grain to rock sample scale. These processes are dependent on local conditions and often coupled. This local and multi-physics nature of the processes controlling the rheology evolution hinders their explicit modelling. An alternative is to focus on a specific process, like grain size reduction by dynamic recrystallization, which can be represented by variation of the fluidity. However, this requires extrapolating empirically-derived flow and grain size evolution laws across orders of magnitude in space and time and conducting tailored simulations for each rock type and deformation conditions (Bercovici et al., 2015; Rozel et al., 2011; Ruh et al., 2024), while potentially overlooking critical feedbacks with other processes. Indeed, dynamic recrystallization also produces softening by eliminating dislocation entanglements that hinder dislocation glide and creating grains with orientations more favorable for glide. These phenomena may be modeled by a decrease of the Peierls stress. Re-deformation of the recrystallized grains will produce increase of the dislocation density and hence of the Peierls stress (healing). To bypass this obstacle, we choose to explore arbitrarily the potential damage intensity space ($\pi_{dam}$) for plausible variations of the Peierls stress and fluidity and model the mechanical consequences of these evolutions. The stochastic formulation of the rheology allows the representation over time of the spatial variability of the evolution.

The potential power for damage ($\pi_{\dot{w}}$) expresses the balance between the rate of energy production in the system and that controlling the change of the material properties. This adimensional parameter implicitly represents the impact of the boundary conditions or changes of the physico-chemical conditions on the evolution of the rheology and, hence, on strain localization. If these conditions are fixed, the rock microstructure will evolve through a given assemblage of processes and the energy cost for changing the rock microstructure will be constant. In nature, however, physico-chemical conditions often vary. An increase in the external forces, leading to higher bulk stresses, or stress concentrations due to the large-scale geometry of the system (e.g., the presence of rigid domains like a craton) or interactions with seismogenic faults will result in an increase in the bulk work-rate in the system, which may "move" a rock volume from the non-damage to the strain localization domain in the regime diagram (Fig. 5). Similarly, changes in pressure, temperature or chemical environment will affect both the deformation and structure evolution processes, modifying the potential power for damage $\pi_{\dot{w}}$.

The similarity of the regime diagrams for strain localization and bulk softening obtained with different constitutive equations and evolution laws (Fig. 7) implies that the main feedbacks triggering and controlling the intensity of viscous strain localization are captured by this simple formulation, with no need for full modelling of the microscale processes modifying the rheology. The present results are therefore also relevant for modelling viscous strain localization in a large variety of non-geological materials, like ice, metals, ceramics, polymers, or complex fluids. However, despite the power of the present approach for unravelling the feedbacks and main parameters controlling strain localization, practical use of the regime diagram is hindered by the difficulty in constraining $\pi_{\dot{W}}$.

To start, the actual partitioning of the mechanical energy dissipation between shear heating and microstructural evolution is difficult to establish (Holtzman et al., 2018; Nassar, 2025); experimental estimates vary widely. In this study, we neglected shear heating. Considering that only a fraction of the work-rate contributes to damage will not change the shape of the regime diagrams, but translate the domains' limits by increasing the mechanical work-rate required to overcome the threshold for damage. Shear heating in itself, at first, enhances strain localization, but thermal diffusion counteracts this effect (Fleitout and Froidevaux, 1980).

The use of the work-rate, rather than the proportion of mechanical work converted to stored energy is another simplification in the present model. However, accounting for stored energy evolution is only possible at the crystal scale for specific deformation processes and conditions. Thus, we assume that steady-state conditions for the energy storage are reached at time scales much shorter than those considered in the models and that the energy available for triggering microstructural evolution may be approached by the local instantaneous work-rate. Moreover, to achieve steady-state before too strong deformation of the mesh, the kinetics of damage and healing were accelerated relative to those observed in experiments (Barnhoorn et al., 2004). This affects the duration of the transient, which has to be rescaled, but has a minor effect on the bulk properties at steady-state.

Finally, the models in the present study were performed under a constant, fixed temperature. Temperature plays a major role on deformation and microstructure evolution processes. By imposing stochasticity and evolution to the Peierls stress ($\tilde{\sigma}$), the model implicitly captures the stronger strain localization observed in nature under moderate-temperature, high-stress conditions. As $\tilde{\sigma}$ modulates the temperature and stress dependences in the constitutive equation, at high temperature, changes in $\tilde{\sigma}$ produce lower effective viscosity contrasts. Consequently, the strain localization domain in the regime diagram is translated towards higher $\pi_{dam}$ values with increasing temperature. However, the effects of temperature are multiple. Higher temperatures lower the strength of rocks, decreasing work-rates ($\pi_{\dot{W}}$). The strength contrasts between minerals and the viscoplastic anisotropy are also reduced at high temperatures (Durham and Goetze, 1977; Holyoke and Tullis, 2006), lowering $\pi_{sto}$. By accelerating diffusion, temperature also changes the damage and healing kinetics. It accelerates recovery and recrystallization, favoring both nucleation and grain growth. Higher temperatures result in more extensive recrystallization and larger recrystallized grain sizes (Barnhoorn et al., 2004; Poirier, 1985). The temperature dependence of all these processes needs to be considered for fully accounting for the effect of temperature on strain localization, but qualitatively all add up to decrease the probability and intensity of strain localization at high temperature.

### *6.2 Comparison to nature and experiments*

By enabling changes in the mechanical behavior as a function of the ratio between the local viscous dissipation rate and the power required to change the rock structure and produce softening and explicitly considering stochasticity, the present models reproduce many key features of ductile deformation in geological materials. First, that strain localization is a stable mode of ductile deformation, which may be, but is not necessarily, triggered by the ubiquitous heterogeneity of rocks. Analysis of natural systems and rock deformation experiments indeed documents that ductile deformation is always heterogeneous at the grain scale, but strain localization at larger scales, characterized by the development of shear zones is not systematic. For instance, high-temperature shear experiments in monomineralic aggregates with fine grain sizes usually display bulk softening in response to evolution of the microstructure, but no strain localization at the sample scale (Bystricky et al., 1999; Hansen et al., 2014; Pieri et al., 1999). In contrast, deformation under low-temperature, high-stress conditions, which is usually accompanied by marked microstructural evolutions, in particular grain size reduction due to dynamic recrystallization, is usually associated with strong strain localization (e.g., Carreras, 2001; Simpson and Schmid, 1983; Vauchez, 1987; White et al., 1980).

These observations are consistent with other outputs of this study. First, triggering of strain localization requires and depends on the intensity of the rheological heterogeneity, but, also on the energy availability for the material properties to evolve. Second, once the rheology starts to evolve, the main factor controlling the intensity of strain localization is the contrast in effective viscosity produced by this evolution. Direct correlation between the evolution of the strength contrast between different mineral phases, strain localization, and bulk mechanical softening in polymineralic rocks subjected to shear has also been documented experimentally (Holyoke and Tullis, 2006). Experiments also document the role of the initial state: under similar conditions, coarse-grained natural samples develop strong strain localization and soften in response to dynamic recrystallization, whereas synthetic fine-grained samples deform homogeneously (Holyoke et al., 2014).

For all three constitutive equations tested, stronger effective viscosity contrasts produce thinner shear zones, which accommodate most of the imposed deformation (Fig. 7). The effective viscosity contrast depends first-order on the damage intensity ($\pi_{dam}$). However, it is modulated by the constitutive equation, which represents the active deformation processes and the dependence of deformation on the physico-chemical conditions, and, for the non-linear rheologies, the imposed strain rates or stresses. The extreme variability of thicknesses of natural ductile shear zones at any scale may therefore result from variations of all these parameters.

### *6.3 Implications for geodynamical modelling*

The present model has no intrinsic length scale. The results can, in principle, be generalized across scales. However, the assumption of an initially random rheological heterogeneity field is only justified at the rock (polycrystal) scale. At larger scales, rheological heterogeneity is common, but neither randomly nor periodically distributed. The constitutive equations and the damage evolution formulation used in this study are also inspired from observations at μm to cm scales. This model is therefore best suited for describing strain localization in rock volumes of $cm^3$ to tens of $m^3$ composed of (initially) randomly arranged grains (or grains ensembles) with contrasted rheological properties.

Modelling strain localization at larger scales requires coarse-graining of the present results. The key point for performing this coarse-graining are the observations that structural evolution (damage) leading to softening at the outcrop and larger scales is predominantly characterized by formation of shear zones (damage and softening confined to a planar volume of finite width), rather than being homogeneous over the entire rock volume. The present study shows that strain localization results in directional softening, whereas damage of the entire volume produces a stronger, but isotropic softening. By neglecting the intermediate-scales and directly jumping from the mm to the km scale in geodynamical models, one overlooks, therefore, an emergent mechanical property – the viscous anisotropy that arises from strain localization. This may represent a significant limitation, as viscous anisotropy - regardless of its origin - has been shown to play a pivotal role on the deformation of lithospheric plates (Duretz et al., 2025; Mameri et al., 2021; Tommasi and Vauchez, 2001).

To model the processes producing strain localization at scales larger than the outcrop, one needs relations quantifying: (1) the probability of a mesh element in a large-scale model to be subjected to strain localization, full damage, or no damage and (2) the local isotropic (if full-damage) or anisotropic (if strain localization) softening relative to the initial non-damaged flow law. Such relations can be parameterized using the present models.

## 6. Conclusion

The adoption of a stochastic description of the rheology, which evolves in time and space as a function of the local mechanical energy dissipation rate, enables to model spontaneous generation of viscous strain localization by self-organization of local heterogeneity into system-scale shear zones. It generates, as an emergent property, viscous anisotropy at the system-scale. By exploring a large parameter range, we document that:

- Strain localization is a steady-state mode of deformation of viscous materials. It results in stable bulk properties (intensity of strain localization, viscosity contrasts, and bulk softening) even when the system is periodically perturbed.
- When the evolution of the rheology results from changes in the rock structure (damage), the rate of energy consumption by the process producing this evolution should lie within a range of the bulk mechanical energy dissipation in the system; this range increases with increasing heterogeneity in the rheology. Otherwise the rock volume is subjected to either full or no damage.
- Because the softening (reduction in effective viscosity) resulting from a change in rock structure is finite, viscous strain localization produces shear zones with finite widths, which depend on the system properties and not on the mesh size.

- The development of system-scale strain localization zones results in effective bulk softening. This softening is stronger than that associated with spatially random distribution of damage, which may be modelled by averaging the mechanical properties, but weaker than that associated with damage of the entire volume.
- Moreover, even when the microscale rheology is isotropic, strain localization results in an anisotropic system-scale mechanical behavior: it strongly reduces the bulk shear viscosity parallel to the shear zones. Softening is minimal at 45° to them.

The present models not only reproduce most qualitative observations in natural and experimental shear zones, but quantitatively predict: (1) whether ductile strain localization, in the form of 'system-scale' shear zones, may or not develop spontaneously in a rock volume, (2) the intensity of strain localization, and (3) the resulting bulk softening as a function of a few characteristics of the system. These characteristics are: (1) the constitutive equation, which represents the active deformation processes, (2) the evolution law for the rock mechanical properties, (3) the initial properties, and (4) the energy input to the system (boundary conditions). This allows the construction of a regime diagram for strain localization based on three adimensional parameters: $\pi_{dam}$, $\pi_{\dot{w}}$, and $\pi_{sto}$.

These results are key for self-consistent modelling of ductile strain localization. The present approach cannot be directly applied to model strain localization in regional and global geodynamical models, but it provides the necessary data for parameterizing the damage and constitutive equations needed to model the effect of viscous strain localization at the sample and outcrop scales – notably the emergence of a directional softening, leading to an anisotropic viscosity at the larger scales. Based on the observation that structural evolution at the outcrop and larger scales produces either strain localization or softening of the entire volume, as modelled in the present study, we posit that modelling of strain localization in geodynamical models requires equations quantifying: (1) the probability of a mesh element to be subjected to strain localization, full damage, or no damage and (2) the isotropic (if full-damage) or anisotropic (if strain localization) softening within the element relative to the initial non-damaged flow law. The model ensembles presented here allow parameterizing these equations. The similarity between the regime diagrams and bulk softening landscapes obtained using different constitutive equations denotes that their form does not depend on the specific processes controlling the rheology and its evolution at the microscale. The present study provides therefore a path for self-consistent modelling the generation and evolution of large-scale shear zones and plate tectonics.

**Funding:** This work was supported by the European Research Council (ERC) under the European Union Horizon 2020 Research and Innovation programme [grant agreement No 882450 – ERC RhEoVOLUTION].

**Acknowledgements**: This work benefited from discussions within the RhEOVOLUTION team during the last 4 years. Special thanks to Jean Michel Brankart for guiding our first steps in stochastic modelling of physical processes and Catherine Thoraval and Alain Vauchez for playing the devil's advocate for many of the ideas exposed in this study. Octave Koechlin and Ulysse Koechlin are thanked for their contribution on parametric analyses in the initial stages of the study, Nestor Cerpa, for constructive criticism on different versions of the manuscript, and Anne Delplanque, for assistance with figures 1 and 3. Laurent Montési and Yannick Ricard are thanked for the constructive/challenging reviews.

**Authors contribution statement: Andréa Tommasi**: Conceptualization, Methodology, Investigation, Writing - Original draft preparation, Project administration, Funding acquisition. **Felipe Sáez-Leiva**: Investigation, Visualization. **Lilou Zeller**: Software, Methodology, Data curation. **Michel Peyret**: Methodology, Software, Investigation. **Riad Hassani**: Software, Validation. **Maurine Montagnat**: Participation to the conceptualization. **All authors**: Writing - Review & Editing

The code ADELI in which the stochastic evolutive rheologies were implemented and the associated documentation are available at https://gitlab.oca.eu/hassani/adeli3p9. Permanent Software Heritage Snapshot to the version used in this study : swh:1:snp:9ea00298cc9096c8cfa5cb3828aec62bc438bfc0. Examples of the input files and csv files summarizing the results of the simulations used for building the regime diagrams are available at https://github.com/Lilou-Zeller/RheoData (https://doi.org/10.5281/zenodo.19481563).

**9. References**

Austin, N.J., Evans, B., 2007. Paleowattmeters: A scaling relation for dynamically recrystallized grain size. Geology 35, 343–346.

Barnhoorn, A., Bystricky, M., Burlini, L., Kunze, K., 2004. The role of recrystallisation on the deformation behaviour of calcite rocks: large strain torsion experiments on Carrara marble. J. Struct. Geol. 26, 885–903.

Barnhoorn, A., Bystricky, M., Kunze, K., Burlini, L., Burg, J.-P., 2005. Strain localisation in bimineralic rocks: Experimental deformation of synthetic calcite-anhydrite aggregates. Earth Planet. Sci. Lett. 240, 748–763.

Bercovici, D., Ricard, Y., 2014. Plate tectonics, damage and inheritance. Nature 508, 513–516.

Bercovici, D., Ricard, Y., 2005. Tectonic plate generation and two-phase damage: void growth versus grainsize reduction. J. Geophys. Res. 110, B03401, doi:10.1029/2004JB003181.

Bercovici, D., Tackley, P., Ricard, Y., Schubert, G., 2015. The generation of plate tectonics from mantle dynamics, in: Treatise on Geophysics. Elsevier, pp. 271–318. https://doi.org/10.1016/B978-0-444-53802-4.00135-4

Boissonneau, G., Tommasi, A., Barou, F., M.A., L.-S., Montagnat, M., 2025. Dynamic recrystallization and mechanical behavior of Mg alloy AZ31: Constraints from tensile tests with in-situ EBSD analysis. Comptes Rendus L Acad. Des Sci. Ser. II Fasc. B - Mec. Phys. Chim. Astron. 353, 235–258. https://doi.org/10.5802/crmeca.267

Braun, J., Chéry, J., Poliakov, A., Mainprice, D., Vauchez, A., Tommasi, A., Daignières, M., 1999. A simple parameterization of strain localization in the ductile regime. J. Geophys. Res. 104, 25,167-25,182.

Brown, M., Solar, G.S., 1998. Shear-zone systems and melts: feedback relations and self-organization in orogenic belts. J. Struct. Geol. 20, 211–227.

Bystricky, M., Burlini, L., Kunze, K., Burg, J.P., 1999. Experimental deformation of olivine aggregates to high strains in torsion. Göttinger Arb. Geol. Paläont. Sb4, 22–23.

Carreras, J., 2001. Zooming on Northern Cap de Creus shear zones. J. Struct. Geol. 23, 1457–1486.

Chardelin, M., Tommasi, A., Padrón-Navarta, J.A., 2024. Progressive Strain Localization and Fluid Focusing in Mantle Shear Zones during Rifting: Petrostructural Constraints from the Zabargad Peridotites, Red Sea. J. Petrol. 65, egae081.
Dabrowski, M., Schmid, D.W., Podladchikov, Y.Y., 2012. A two-phase composite in simple shear: Effective mechanical anisotropy development and localization potential. J. Geophys. Res. 117, B08406, doi:10.1029/2012JB009183.
Doukhan, J.C., Trépied, L., 1985. Plastic deformation of quartz single crystals. Bull. Minéralogie 108, 97–123.
Drury, M.R., Vissers, R.L.M., Van der Wal, D., Strating, E.H.H., 1991. Shear localisation in upper mantle peridotites. Pure Appl. Geophys. 137, 439–460.
Duretz, T., Petri, B., Mohn, G., Schmalholz, S.M., Schenker, F.L., Müntener, O., 2016. The importance of structural softening for the evolution and architecture of passive margins. Sci. Rep. 6, 38704. https://doi.org/10.1038/srep38704
Duretz, T., Schmalholz, S.M., Kulakov, R., Mohn, G., Tugend, J., Halter, W., Bardroff, A., 2025. Lithospheric Deformation With Mechanical Anisotropy: A Numerical Model and Application to Continental Rifting. Geochemistry, Geophys. Geosystems 26. https://doi.org/10.1029/2025GC012409
Durham, W.B., Goetze, G., 1977. Plastic flow of oriented single crystals of olivine. 1. Mechanical data. J. Geophys. Res. 82, 5737–5753.
Fleitout, L., Froidevaux, C., 1980. Thermal and mechanical evolution of shear zones. J. Struct. Geol. 2, 159–164.
Gardner, R., Piazolo, S., Evans, L., Daczko, N., 2017. Patterns of strain localization in heterogeneous, polycrystalline rocks–a numerical perspective. Earth Planet. Sci. Lett. 463, 253–265.
Gerbi, C., Johnson, S.E., Shulman, D., Klepeis, K., 2016. Influence of microscale weak zones on bulk strength. Geochemistry, Geophys. Geosystems 17, 4064–4077. https://doi.org/10.1002/2016GC006551
Gouriet, K., Cordier, P., Garel, F., Thoraval, C., Demouchy, S., Tommasi, A., Carrez, P., 2019. Dislocation dynamics modelling of the power-law breakdown in olivine single crystals: Toward a unified creep law for the upper mantle. Earth Planet. Sci. Lett. 506, 282–291.
Handy, M.R., 1994. The energetics of steady-state heterogeneous shear in mylonitic rock. Mater. Sci. Eng. A175, 261–272.
Hansen, L.N., Zhao, Y.H., Zimmerman, M.E., Kohlstedt, D.L., 2014. Protracted fabric evolution in olivine: Implications for the relationship among strain, crystallographic fabric, and seismic anisotropy. Earth Planet. Sci. Lett. 387, 157–168.
Higgie, K., Tommasi, A., 2012. Feedbacks between deformation and melt distribution in the crust–mantle transition zone of the Oman ophiolite. Earth Planet. Sci. Lett. 359, 61–72.
Holtzman, B.K., Chrysochoos, A., Daridon, L., 2018. A Thermomechanical Framework for Analysis of Microstructural Evolution: Application to Olivine Rocks at High Temperature. J. Geophys. Res. Solid Earth 123, 8474–8507. https://doi.org/10.1029/2018JB015613
Holyoke, C.W., Kronenberg, A.K., Newman, J., 2014. Microstructural evolution during strain localization in dolomite aggregates. J. Struct. Geol. 69, 449–464.
Holyoke, C.W., Tullis, J., 2006. Mechanisms of weak phase interconnection and the effects of phase strength contrast on fabric development. J. Struct. Geol. 28, 621–640.
Katz, R.F., Spiegelman, M., Holtzman, B., 2006. The dynamics of melt and shear localization in partially molten aggregates. Nature 442, 676–679.
Mameri, L., Tommasi, A., Signorelli, J., Hassani, R., 2021. Modelling olivine-induced viscous anisotropy in fossil mantle strike-slip shear zones and the resulting strain localization in the crust. Geophys. J. Int. 224, 608–625. https://doi.org/10.1093/gji/ggaa400

Meyer, S.E., Kaus, B.J.P., Passchier, C., 2017. Development of branching brittle and ductile shear zones: A numerical study. Geochemistry, Geophys. Geosystems 18, 2054–2075. https://doi.org/10.1002/2016GC006793
Montési, L.G.J., 2013. Fabric development as the key for forming ductile shear zones and enabling plate tectonics. J. Struct. Geol. 50, 254–266. https://doi.org/10.1016/j.jsg.2012.12.011
Nassar, O., 2025. On the conversion of plastic work into heat in metals. Proc. R. Soc. A Math. Phys. Eng. Sci. 481, 202401139. https://doi.org/10.1098/rspa.2024.0139
Neves, S.P., Tommasi, A., Vauchez, A., Carrino, T.A., 2021. The Borborema strike-slip shear zone system (NE Brazil): Large-scale intracontinental strain localization in a heterogeneous plate. *Lithosphere*, 2021(6): 6407232. https://doi.org/10.2113/2021/6407232
Newman, J., Lamb, W.M., Drury, M.R., Vissers, R.L.., 1999. Deformation processes in a peridotite shear zone: reaction-softening by an H2O-deficient, continuous net transfer reaction. Tectonophysics 303, 193–222. https://doi.org/10.1016/S0040-1951(98)00259-5
Pieri, M., Kunze, K., Burlini, L., Stretton, I., Olgaard, D., Burg, J.P., 1999. Experimental deformation to large shear strain of calcite rocks: rheological and microstructural evolution. Göttinger Arb. Geol. Paläont. Sb4, 154–155.
Poirier, J.P., 1985. Creep of crystals. High temperature deformation processes in metals, ceramics and minerals. Cambridge University Press, Cambridge.
Rozel, A., Ricard, Y., Bercovici, D., 2011. A thermodynamically self-consistent damage equation for grain size evolution during dynamic recrystallization. Geophys. J. Int. 184, 719–728.
Ruh, J., Behr, W.M., Tokle, L., 2024. Effect of grain-size and textural weakening in polyphase crustal and mantle lithospheric shear zones. Tektonika 2, 91–110. https://doi.org/10.55575/tektonika2024.68.1.2
Simpson, C., Schmid, S.M., 1983. An evaluation of criteria to deduce the sense of movement in sheared rocks. Geol. Soc. Am. Bull. 94, 1281–1288.
Tommasi, A., Vauchez, A., 2015. Heterogeneity and anisotropy in the lithospheric mantle. Tectonophysics 661, 11–37.
Tommasi, A., Vauchez, A., 2001. Continental rifting parallel to ancient collisional belts: An effect of the mechanical anisotropy of the lithospheric mantle. Earth Planet. Sci. Lett. 185, 199–210.
Tommasi, A., Vauchez, A., Daudré, B., 1995. Initiation and propagation of shear zones in a heterogeneous continental lithosphere. J. Geophys. Res. 100, 22,22-83,101.
Tommasi, A., Vauchez, A., Fernandes, L.A.D., Porcher, C.C., 1994. Magma-assisted strain localization in an orogen-parallel transcurrent shear zone of southern Brazil. Tectonics 13, 421–437.
Urai, J.L., Means, W.D., Lister, G.S., 1986. Dynamic recrystallization of minerals, in: Hobbs, B.E., Heard, H.C. (Eds.), Mineral and Rock Deformation: Laboratory Studies. American Geophysical Union, Washington, DC, pp. 161–199.
Vauchez, A., 1987. The development of discrete shear-zones in a granite: stress, strain and changes in deformation mechanisms. Tectonophysics 133, 137–156.
Vilotte, J.-P., Daignières, M., Madariaga, R., 1982. Numerical modeling of intraplate deformation: simple mechanical models of continental collision. J. Geophys. Res. 87, 10709–10728.
White, S.H., Burrows, S.E., Carreras, J., Shaw, N.D., Humphreys, F.J., 1980. On mylonites in ductile shear zones. J. Struct. Geol. 2, 175–187.

**Supplementary Material for ' Modelling ductile strain localization with evolutive stochastic rheologies' is composed of:**

The presentation of the implementation of the evolutive stochastic rheology.

Table S1 presenting the material properties used in the simulations

Fig. S1 showing $\pi_{\dot{w}} - \pi_{dam}$ sections at different fixed $\pi_{sto}$ of the regime diagram for strain localization in Fig. 4 of the main text.

Fig. S2 showing that the steady-state strain localization is almost independent on the characteristic length scale of the initial rheological heterogeneity field.

Fig. S3 showing that the steady-state strain localization is independent on the mesh refinement.

Fig. S4 illustrating the effect of varying the kinetics of damage and healing on the strain localization regime diagram.

Fig. S5 illustrating the effect of varying the kinetics of damage and healing on evolution of the system through time.

Fig. S6 documenting the softening by the evolution of the bulk stress (Seq) through time in simulations where stochasticity and evolution are imposed on the Peierls stress for different kinetics of the evolution of the rheology.

Fig. S7 documenting the softening by the evolution of the bulk stress (Seq) through time in simulations where stochasticity and evolution are imposed on the Peierls stress for different stochasticity levels

Fig S8 showing examples for pure shear boundary conditions.

# Supplementary material : Formulation of the stochastic evolving rheology

The stochastic material parameter in the flow law is expressed as $P = \overline{P}+P'$ where $\overline{P}$ designates the mean of the local gaussian distribution and $P'$ designates the perturbation.

## 1 Evolution of the rheology

Damage and healing are implemented as a linear evolution of a chosen parameter $\overline{P}$ (in this study, either Peierls stress $\tilde{\sigma}$ or fluidity $\gamma$) between two end-member values : one for the initial undamaged rheology $\overline{P}_0$, and one for the fully damaged one $\overline{P}_{dam}$. In each mesh cell, $\overline{P}$ evolves linearly according to $\Delta J_2(\varepsilon)$ the local deviatoric strain accumulated since the last rheological update :

If $\dot{W}_{cell} > \dot{W}_{th}$ then

$$\begin{cases} \overline{P}_{new} = \overline{P}_{old} + \dfrac{\overline{P}_0 - \overline{P}_{dam}}{K_{dam}} \Delta J_2(\varepsilon) & \text{if } \overline{P}_{new} \in [\overline{P}_{dam}; \overline{P}_0] \\ \overline{P}_{new} = \overline{P}_{dam} & \text{otherwise} \end{cases}$$

else

$$\begin{cases} \overline{P}_{new} = \overline{P}_{old} + \dfrac{\overline{P}_{dam} - \overline{P}_0}{K_{heal}} \Delta J_2(\varepsilon) & \text{if } \overline{P}_{new} \in [\overline{P}_{dam}; \overline{P}_0] \\ \overline{P}_{new} = \overline{P}_0 & \text{otherwise} \end{cases}$$

The results are formulated in terms of $K_{dam}$ the *damage kinetic* which designates the accumulated deviatoric strain required to go from undamaged to fully damaged state (supposing that the work-rate threshold is always met) and in terms of $K_{heal}$ the *healing kinetic* designates the accumulated deviatoric strain required to go from fully damaged state to undamaged state (supposing that the work-rate threshold is never met).

In the present study, if not otherwise stated, $K_{dam} = K_{heal} = 0.01$ which means that if the condition for damage is always met, a cell is completely damaged after an accumulated strain of $\Delta J_2(\varepsilon) = 0.01$ and that a fully damaged cell is completely healed after an accumulated strain of $\Delta J_2(\varepsilon) = 0.01$ since the end of damage if the condition for damage is never met during this period.

This formulation of the evolution law is well-adapted to describe the evolution of the Peierls stress. Damage (decrease of the Peierls stress) results from decrease in this dislocation density and grain reorientation due to dynamic recrystallization, whereas healing results from increase in dislocation density within the previously recrystallized grains due to re-deformation. Both processes depend on the finite strain. If the variable used for describing the variability and evolution of the rheology is the fluidity and the variation in fluidity results from grain size evolution, damage is still produced by grain size reduction due to dynamic recrystallization and depends on finite strain, but healing results from grain growth, which depends on time (and temperature) and is most effective under post-dynamic or static conditions. The healing law has therefore to be adapted to correctly describe this process.

## 2 Stochasticity

### 2.1 Gaussian noise

The rheology is made stochastic by adding to the material property field P a Gaussian noise with spatial correlation.

In every cell, a noise value $P'_{\text{no filt}}$ is drawn from a "Gaussian" distribution centered at 0 with a standard deviation of $std = \pi_{sto} \times \overline{P}$.

Both tails of the "Gaussian" distribution are truncated to ensure that $P'$ remains within physically realistic limits, as in Juricke et al. (2013).

## 2.2 Spatial correlation

The unfiltered noise is spatially correlated to obtain heterogeneities that cover several mesh cells. For each cell, this is done by correcting the unfiltered noise in the cell by the weighted average of the noises in the neighboring cells as follows:

$$P'(i) = \frac{P'_{\text{no filt}}(i) + \sum_k q_{lk} P'_{\text{no filt}}(k)}{(1 + \sum_k q_{lk}^2)^{1/2}} \quad \text{with } q_{ik} = e^{\frac{-d_{ik}}{d_{corr}}}$$

where $d_{ik}$ is the distance between the centroids of the mesh elements $i^{\text{th}}$ and $k^{\text{th}}$ and $d_{\text{corr}}$ the spatial correlation length. This formulation allows to preserve the truncated gaussian formulation of previously chosen mean value and standard deviation (Juricke et al., 2013).

A cutoff point is applied at $2 \times d_{corr}$. Only cells within a distance of $2 \times d_{corr}$ are considered.

## 2.3 Convergence condition on rheology update

In Adeli2024, the force balance equation is approximated by a quasi-static (no acceleration) problem. After finite element discretization, the equilibrium between internal nodal forces (Fint) and external nodal forces (Fext) can be expressed as:

$$F_{int} + F_{ext} = 0$$

The problem is solved using dynamic relaxation (Cundall and Board, 1988; Chéry et al., 2001), a method that consists in adding inertia and damping forces to the problem and allowing it to converge toward the quasi-static solution. Thus, the above equation is replaced by

$$F_{int} + F_{ext} = M\ddot{U} + C$$

where $M$ is the numerical mass matrix and $C$ the numerical damping force

$C = \alpha \left| F_{int} + F_{ext} \right| \frac{\dot{U}}{|\dot{U}|}$ with $\dot{U}$ the velocity vector.

The importance of this purely numerical term, which should remain low, is measured by $\sqrt{\frac{||F_{int}+F_{ext}||^2}{||F_{int}||^2+||F_{ext}||^2}}$
This residual, which is monitored so that numerical effects are not interpreted as physical results, tends towards zero with increasing simulation time. However, the updates of the rheology field change the force equilibrium, increasing the residual momentaneously. They have therefore to be spaced in time.
In this study we chose to update the rheology at 0.001 bulk finite strain steps under the condition that the residual is lower than 0.005%. Time steps are chosen small enough so that the convergence condition is rarely not fulfilled. This may occur, however, at the initialization of the models, delaying the first update of the rheology.

Chery J., Zoback M., Hassani R., An integrated mechanical model of the San Andreas Fault in central and northern California, 2001, Solid Earth

Cundall, P.A., and M. Board, A micro computer program for modelling large strain plasticity problems, in Numerical methods in Geomechanics, edited by G. Swoboda, pp. 2101-2108, Balkema, Rotterdam, 1988.

Juricke S., Lemke P., Timmermann R., Rackow T., Effects of Stochastic Ice Strength Perturbation on Arctic Finite Element Sea Ice Modeling, 2013, Journal of Climate, 3785-3802

Table 1: Material parameters used at the initial state of the different sets of numerical experiments.

| Parameter | Unit | Simulation type | | |
|---|---|---|---|---|
| | | "Peierls" | "Power law n=3" | "Newtonian n=1" |
| Young modulus | GPa | 200 | 200 | 200 |
| Poisson coefficient | | 0.25 | 0.25 | 0.25 |
| Fluidity $\boldsymbol{\gamma}$ | $Pa^{-n}\ s^{-1}$ | $2\times10^{11}$ | $\mathbf{3\times10^{-17}}$ | $\mathbf{1\times10^{-3}}$ |
| Activation energy Q | $kJ\ mol^{-1}K^{-1}$ | 460 | 460 | 370 |
| Peierls stress $\tilde{\boldsymbol{\sigma}}$ | GPa | **2** | - | - |
| Stress exponent n | | 3 | 3 | 1 |
| p | | 1.5 | - | - |
| q | | 2 | 0 | 0 |

The parameter to which stochasticity is applied is indicated in bold. The value displayed is the mean of the initial probability distribution function. Values from Gouriet et al. (2019). In the stochastic fluidity simulations, the initial fluidity has been adjusted to obtain initial bulk effective viscosities equivalent (same order of magnitude) to those in the stochastic Peierls stress simulations.

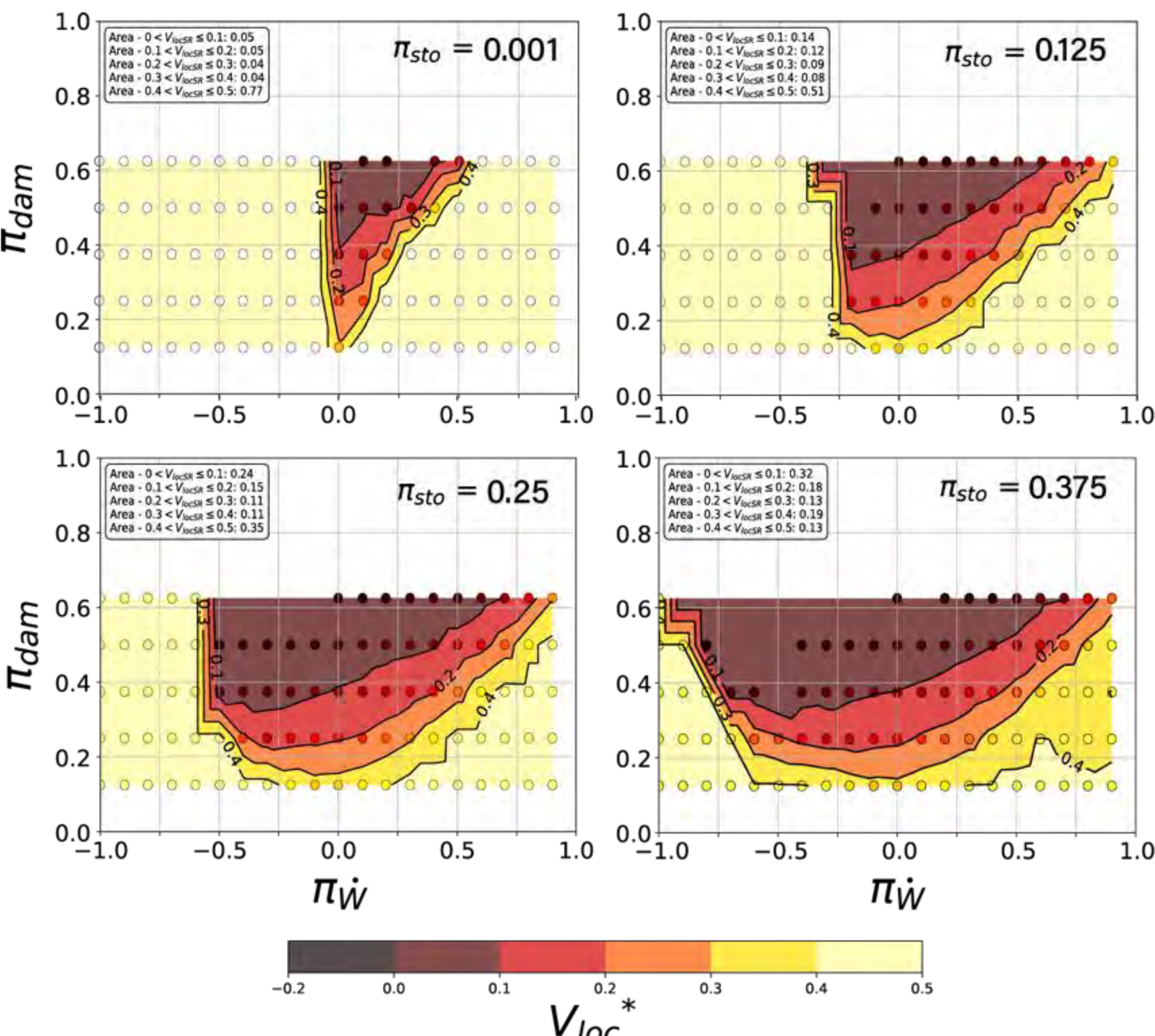

**Fig. S1: Effect of stochasticity on strain localization**. $\pi_{\dot{W}} - \pi_{dam}$ sections at different $\pi_{sto}$ of the regime diagram for viscous strain localization in system where stochasticity and evolution are imposed on the Peierls stress (different standard deviations of the rheological property probability distribution function). In models with very low stochasticity, strain localization is solely observed when the initial bulk work rate is either equal or slightly higher than the threshold. Increasing the spatial variability of the rheology enhances the strain localization domain at the expenses of the no damage one by enhancing the probability for the local work-rate to overcome the threshold and, hence the probability of local damage. It also enhances the strain localization domain at the expenses of the fully damage one by enhancing the spatial variability of the rheology during the transient.

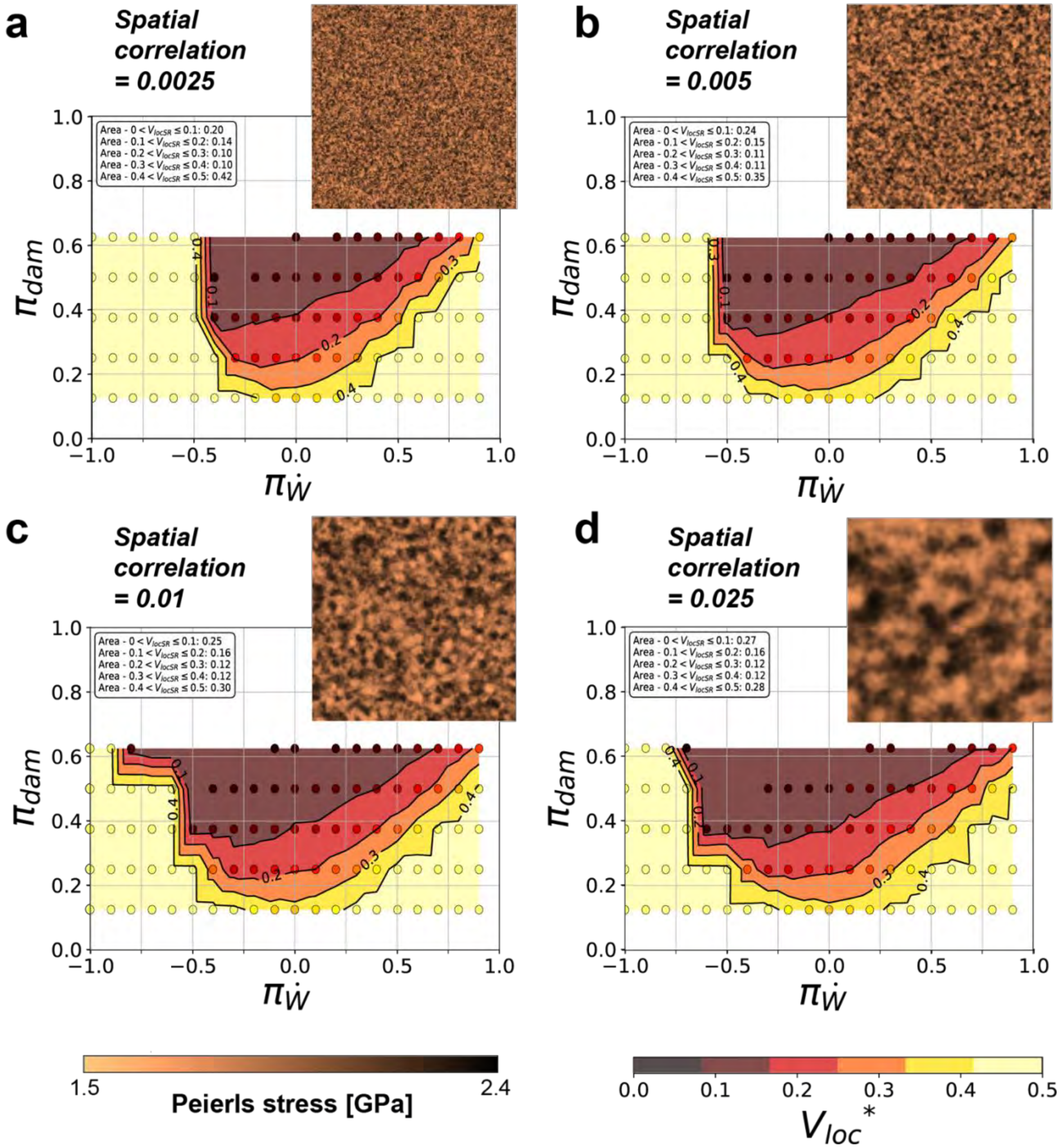

**Fig. S2: Effect of the ratio between the initial characteristic length scale of the heterogeneities and that of the system**. The initial characteristic length scale of the rheological heterogeneity field is varied by changing the spatial correlation imposed on the noise. All simulations have the same mesh, meaning that that the heterogeneities are described by varying number of elements. Even in (a) the heterogeneities are discretized by more than one element to ensure a correct description of their mechanical behavior. The shape of the strain localization domain and the intensity of the strain localization at steady-state are almost independent of the initial heterogeneity field length scale. Increase by a factor 100 on the initial heterogeneity size only slightly enhances the strain localization domain and the intensity of strain localization, due to the dependence of the characteristic length scales of strain rate and stress perturbations associated with each heterogeneity on the heterogeneity size. The perturbation distance d decreases, for the highly non-linear flow law used in these simulations as $\frac{1}{\left({}^{d}/_{r}\right)^{s}}$ with $s$=[1,2] and $r$, the mean heterogeneity radius. Stronger strain localization results in a larger number of simulations in which the mesh degenerates before a bulk strain of 0.1 is achieved (missing points in the regime diagrams). All plots for $\pi_{sto}$ = 0.25.

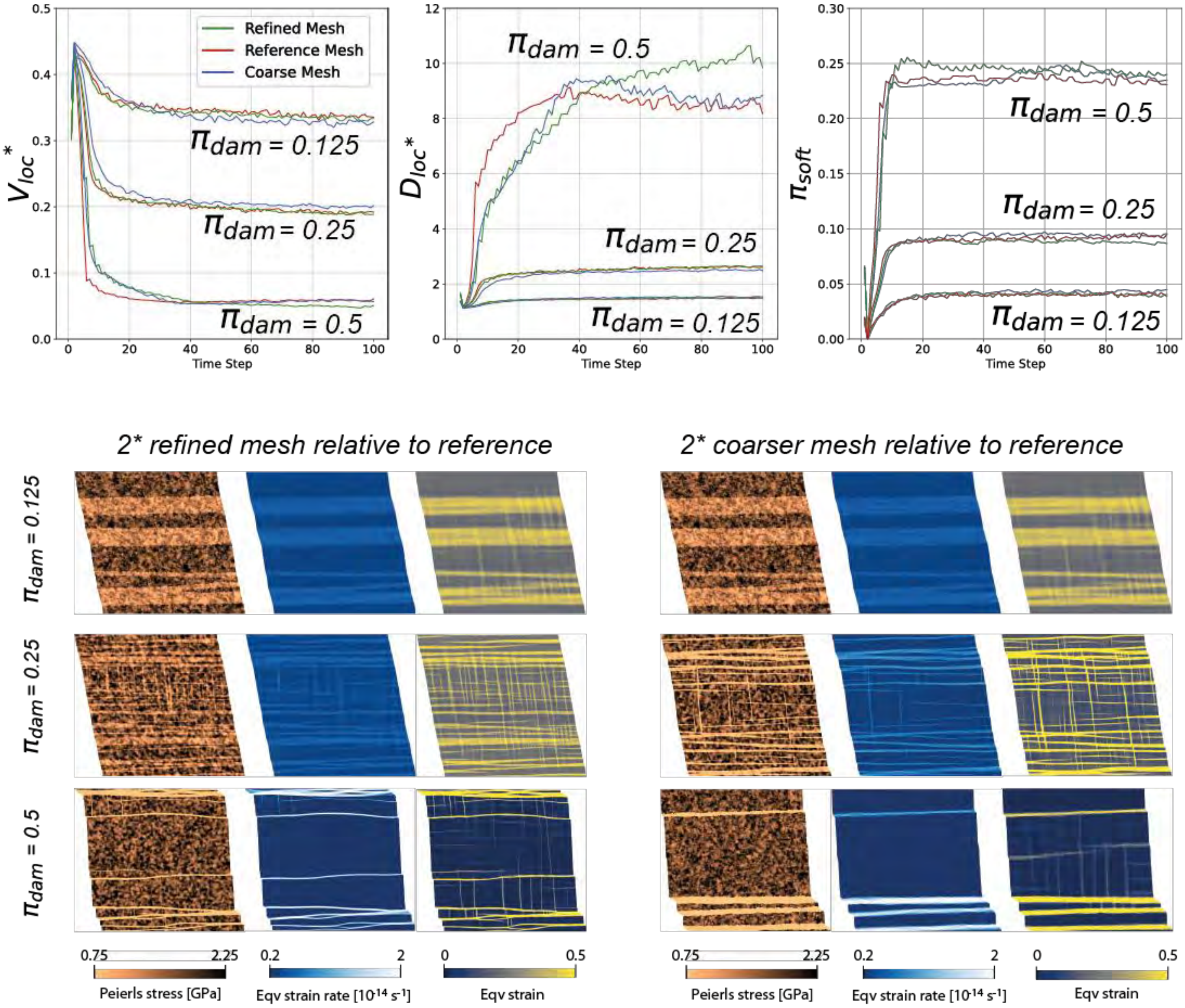

**Fig. S3: Mesh resolution tests,** which show that the steady-state behavior of the system – the volume fraction accommodating the localized deformation ($V_{loc}$*), the intensity of the strain localization ($D_{loc}$*) and the bulk softening ($\pi_{soft}$) - do not depend on the mesh discretization. These results also show that the width of the shear zones does not decrease to a single element with increasing mesh refinement, although a better discretization allows for the expression of more and thinner shear zones in higher $\pi_{dam}$ simulations. A finer mesh also enables a wider sampling of the stochasticity, producing more variability of the strain rate field and delaying the achievement of steady-state.

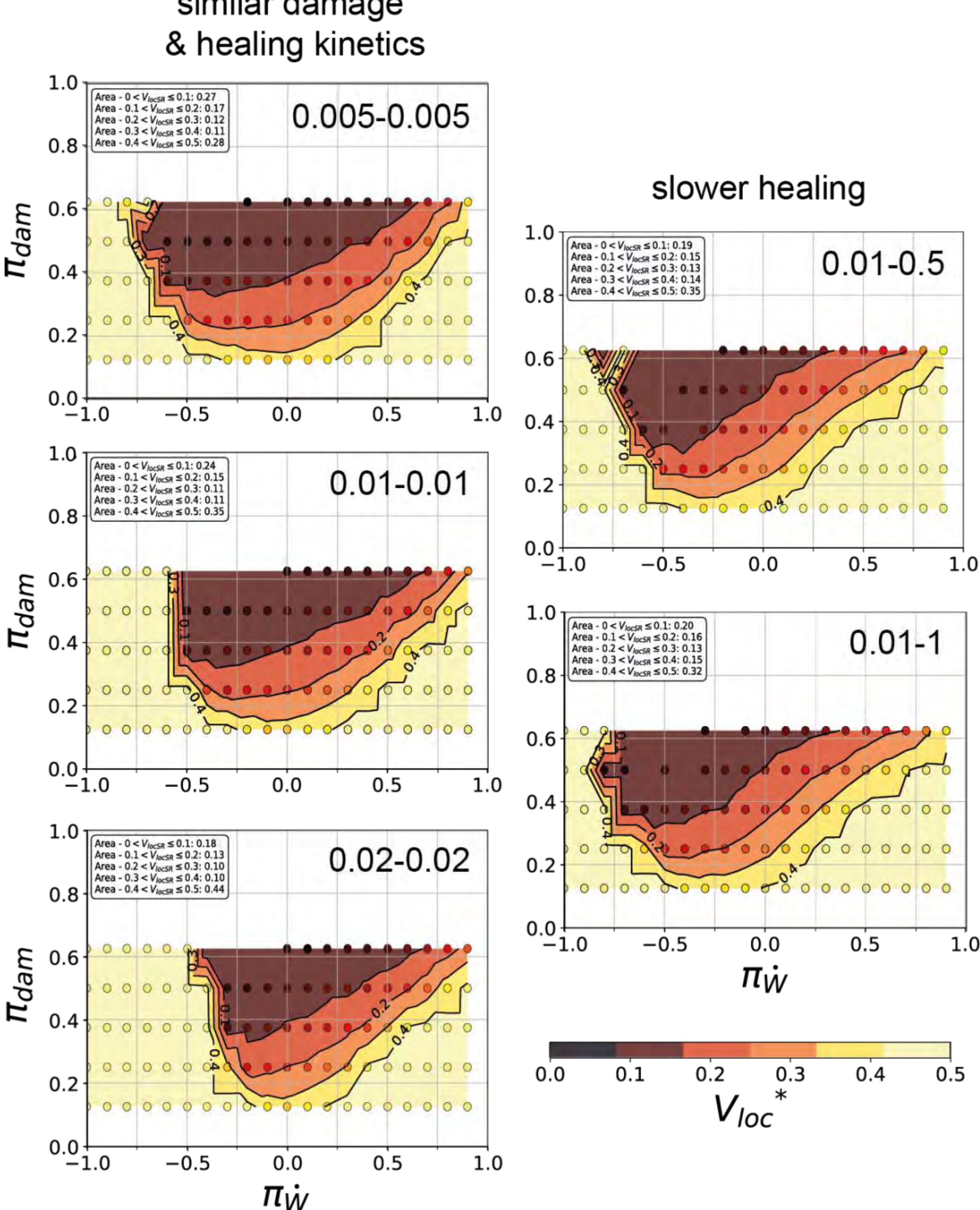

**Fig. S4: Effect of varying the kinetics of damage and healing on the strain localization regime diagram,** implemented by imposing different scaling factors $\left(\frac{\partial P}{\partial \varepsilon}\right)$ in the evolution equation (cf. Fig. 2 in the main text); the first value on the top right of the plots is the damage rate, the second is the healing rate. Fast evolution of the rheology relative to the strain rate slightly enlarges the strain localization domain at the expenses of the no-damage one. Slower healing relative to damage moves the strain localization domain leftwards. As expected, slower healing results in increase of the full damage domain relatively to the strain localization one. It also enhances the strain localization domain at the expenses of the no damage one, as damaged zones have a longer life span, increasing the probability of a new damage event in the vicinity triggering a positive feedback, which will lead to strain localization. These effects are also documented in the evolution of $V_{loc}$* through time (Fig. S5). All plots for $\pi_{sto}$ = 0.25.

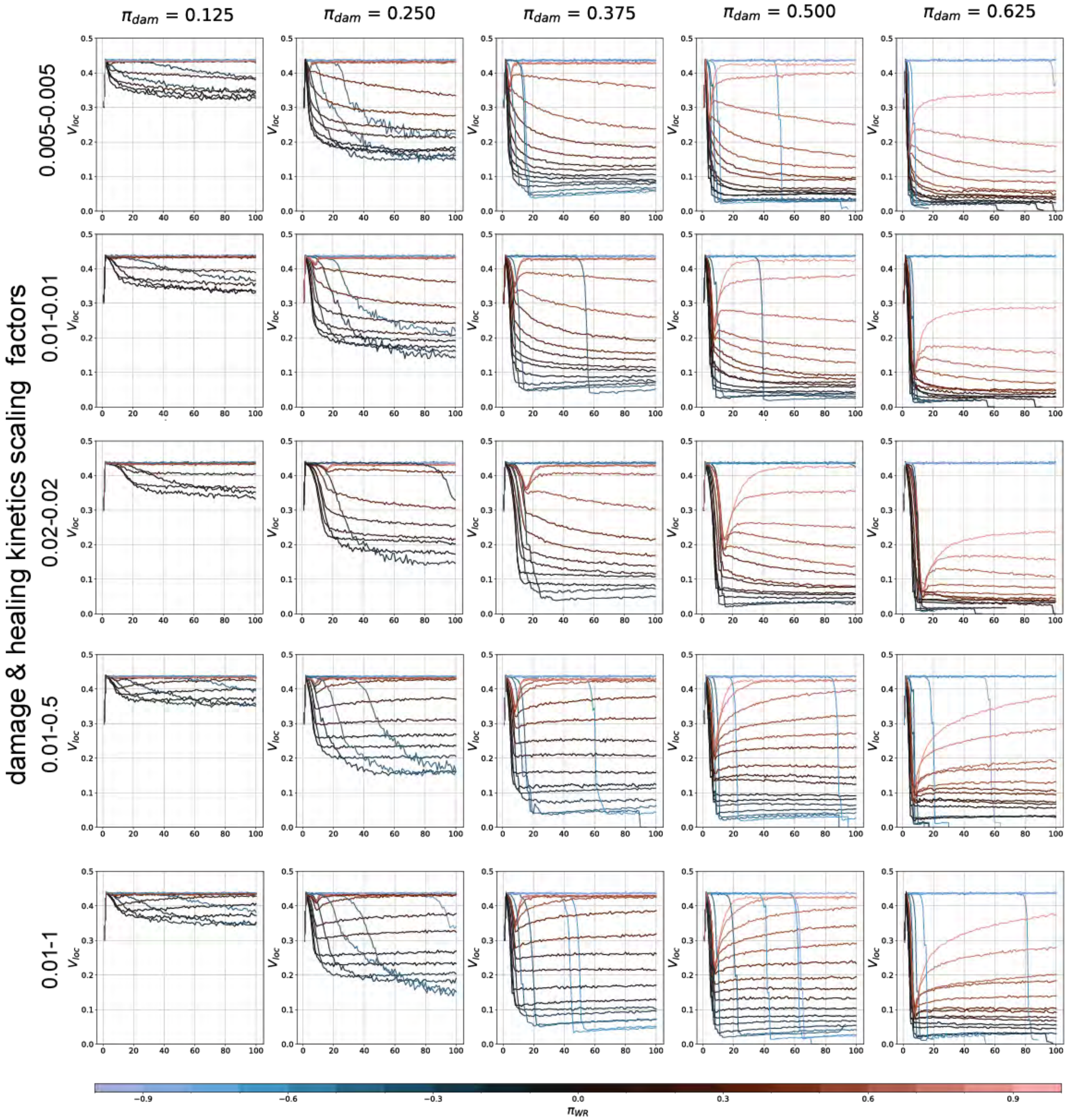

**Fig. S5:** Effect of varying the kinetics of damage and healing on the evolution of strain localization ($V_{loc}$*) through time, in simulations where stochasticity and evolution are imposed on the Peierls stress. All plots for $\pi_{sto}$ = 0.25.

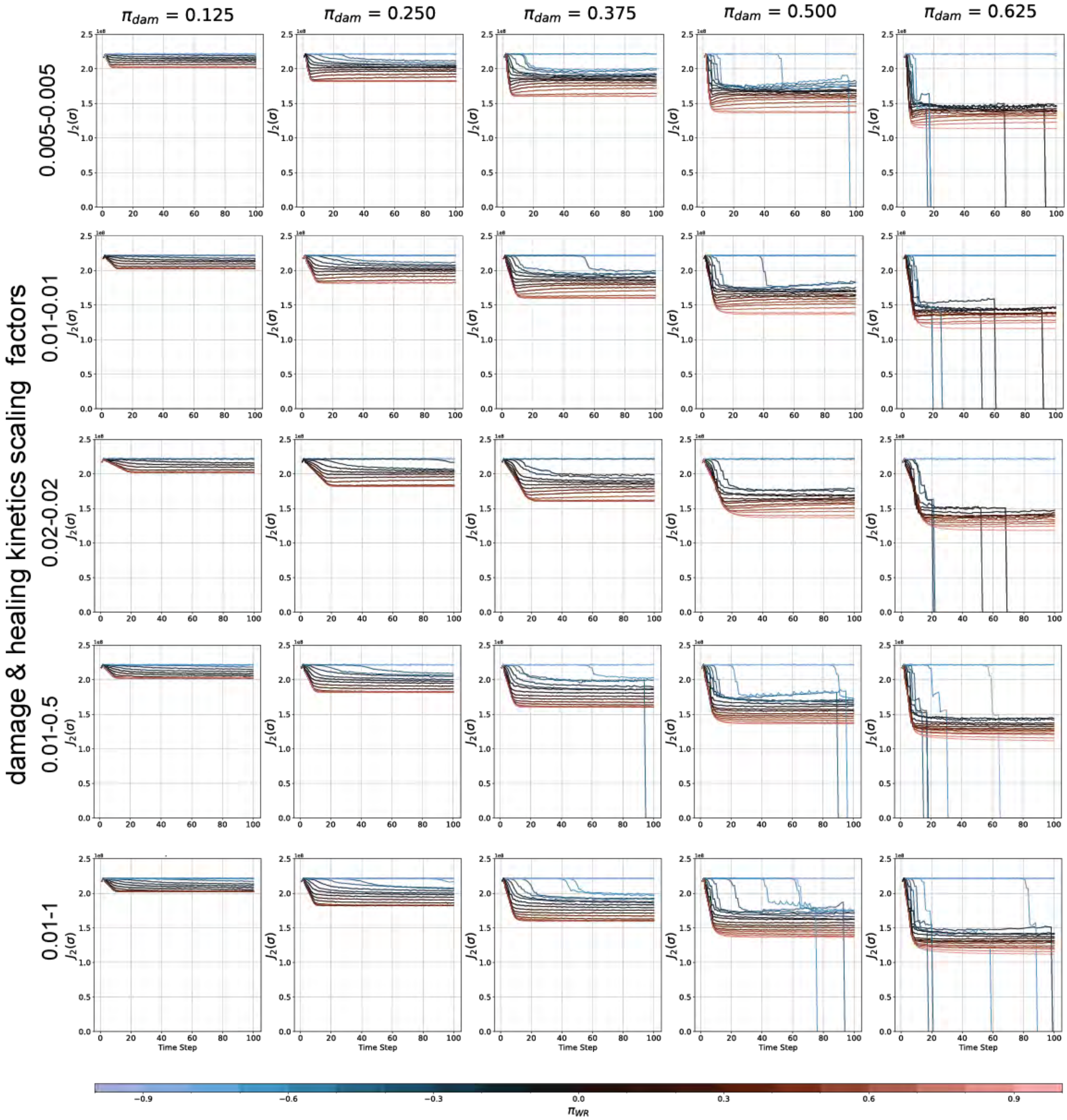

**Fig. S6**: **Bulk softening**, characterized by the evolution of the bulk stress (Seq) through time, in simulations where stochasticity and evolution are imposed on the Peierls stress for different kinetics of the evolution of the rheology. All plots for $\pi_{sto}$ = 0.25.

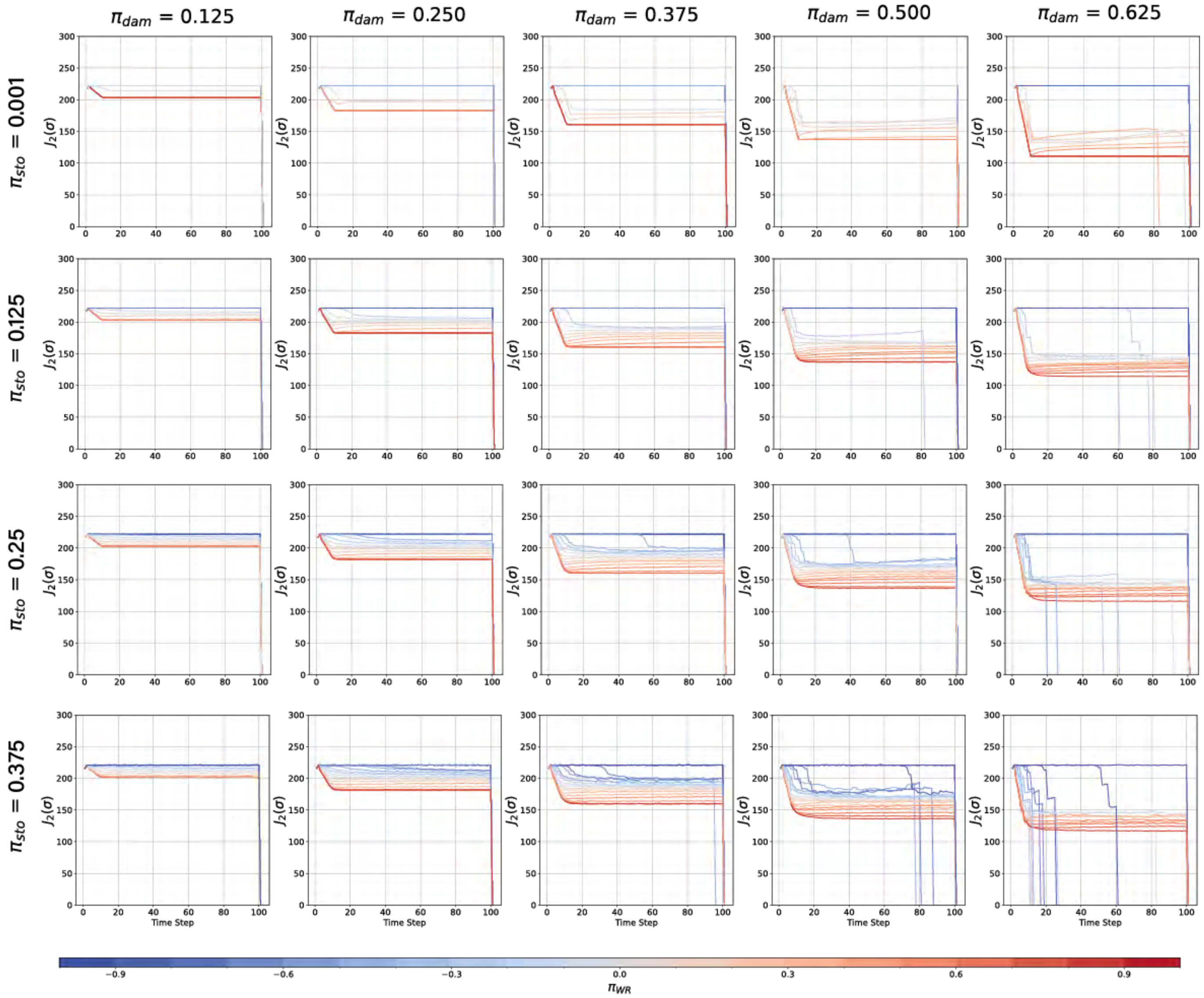

**Fig. S7**: **Bulk softening**, characterized by the evolution of the bulk stress (Seq) through time, in simulations where stochasticity and evolution are imposed on the Peierls stress for different stochasticity degrees. Increasing stochasticity enables strain localization in a larger domain of the parameter space, but has otherwise a minor effect on the bulk softening.

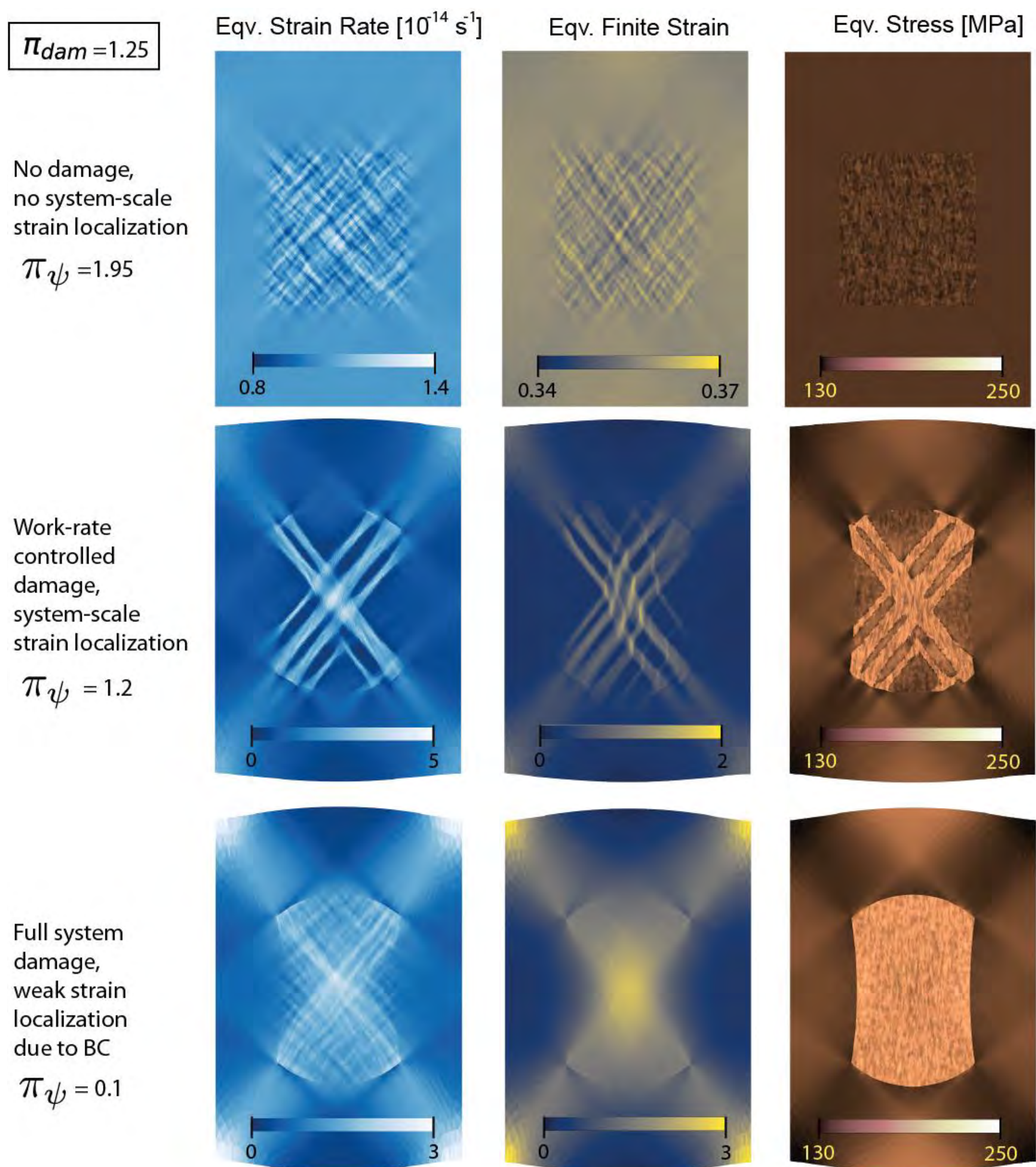

**Fig. S8:** (a) Finite strain and (b) deviatoric stress fields at the end of the experiments for simulations with pure shear boundary conditions (imposed shortening parallel to X, null displacements parallel to Y, free displacements parallel to Z) with potential energy for damage ($\pi_{\psi} = 1$) and potential intensity of damage ($\pi_{dam} = 0.125$)). These simulations display the same evolution paths as the simple shear ones: strain localization or no strain localization with either no or full damage, and corroborate strain localization produces shear zones at 45° to the maximum and minimum deviatoric stresses. The pure shear models are, however, ill-fitted to define regime diagrams because (1) the imposed strain rate varies during the simulation due to the changes in the box length and width, (2) advection rotates the shear zones out of the 'ideal' orientation, and (3), in extension, spatial variations in the box width due to strain localization trigger necking. Moreover, they require either a buffer zone (as in the examples shown here) or very anisometric boxes (length >> width) to avoid reflections of the shear zones at the boundaries on which the shortening or extension is imposed and even with a buffer zone, at higher strains, perturbations due to the boundary conditions are observed, as it can be observed in the figure above, in particular for the full damage experiment.